\documentclass[pdflatex,sn-basic]{sn-jnl}
\theoremstyle{thmstyleone}%
\theoremstyle{thmstyletwo}%

\theoremstyle{thmstylethree}%

\usepackage{multirow}
\begin{document}

\title[HASPIDE Space]{A Hydrogenated amorphous silicon detector for Space Weather Applications}
\author[1,2]{Grimani Catia} \equalcont{These authors contributed equally to this work.}
\author[1,2]{Fabi Michele} \equalcont{These authors contributed equally to this work.}
\author[1,2]{Sabbatini Federico} \equalcont{These authors contributed equally to this work.}
\author[1,2]{Villani Mattia} \equalcont{These authors contributed equally to this work.}
\author[3]{Antognini Luca}
\author[4,5]{Bashiri Aishah}
\author[6,7]{Calcagnile Lucio}
\author[6,7]{Caricato Anna Paola}
\author[8]{Catalano Roberto}
\author[2]{Chil\`a Deborah}
\author[8]{Cirrone Giuseppe Antonio Pablo}
\author[9,10]{Croci Tommaso}
\author[8]{Cuttone Giacomo}
\author[3]{Dunand Sylvain}
\author[11]{Frontini Luca$^{11}$}
\author[9]{Ionica Maria}
\author[12,9]{Kanxheri Keida}
\author[4]{Large Matthew}
\author[11]{Liberali Valentino$^{11}$}
\author[6,7]{Martino Maurizio}
\author[6,7]{Maruccio Giuseppe}
\author[13]{Mazza Giovanni}
\author[9]{Menichelli Mauro}
\author[6,7]{Monteduro Anna Grazia}
\author[9]{Morozzi Arianna}
\author[9,14]{Moscatelli Francesco}
\author[15,2]{Pallotta Stefania}
\author[9,10]{Passeri Daniele}
\author[14,9]{Pedio Maddalena}
\author[4]{Petasecca Marco}
\author[8]{Petringa Giada}
\author[9,12]{Peverini Francesca}
\author[13]{Piccolo Lorenzo}
\author[10,9]{Placidi Pisana}
\author[6,7]{Quarta Gianluca}
\author[6,7]{Rizzato Silvia}
\author[11]{Stabile Alberto$^{11}$}
\author[15,2]{Talamonti Cinzia}
\author[3]{Thomet Jonathan}
\author[9]{Tosti Luca}
\author[13]{Wheadon Richard James}
\author[3]{Wyrsch Nicolas}
\author[9,16] {Zema Nicola}
\author[9]{Servoli Leonello} \equalcont{These authors contributed equally to this work.}

\affil[1]{DiSPeA, Università di Urbino Carlo Bo, Via S. Chiara 27, 61029 Urbino, Italy}
\affil[2]{Istituto Nazionale di Fisica Nucleare  Sezione di Firenze, Via B. Rossi 1, 50019 Sesto Fiorentino, Firenze, Italy}
\affil[3]{Ecole Polytechnique Fédérale de Lausanne (EPFL),Photovoltaic and Thin-Film 
Electronics Laboratory (PV-Lab), Rue de la Maladière 71b, 2000 Neuchâtel, Switzerland}
\affil[4]{Centre for Medical Radiation Physics, University of Wollongong, Northfields Ave Wollongong NSW 2522, Australia}
\affil[5]{Najran University, King Abdulaziz Rd, Najran, Saudi Arabia}
\affil[6]{Istituto Nazionale di Fisica Nucleare  Sezione di Lecce, Via per Arnesano, 73100 Lecce, Italy}
\affil[7]{Department of Mathematics and Physics "Ennio de Giorgi" University of Salento, Via per Arnesano, 73100 Lecce, Italy}
\affil[8]{Istituto Nazionale di Fisica Nucleare  Laboratori Nazionali del Sud, Via S.Sofia  62, 95123 Catania, Italy}
\affil[9]{Istituto Nazionale di Fisica Nucleare (INFN), Sezione di Perugia, Via A. Pascoli snc, 06123 Perugia, Italy}
\affil[10]{University of Perugia, Department of Engineering, Via Goffredo Duranti, 93, 06125 Perugia, Italy}
\affil[11]{Istituto Nazionale di Fisica Nucleare Sezione di Milano, Via Celoria 16, 20133 Milano, Italy}
\affil[12]{University of Perugia, Department of Physics and Geology, Via A. Pascoli snc, 06123 Perugia, Italy}
\affil[13]{Istituto Nazionale di Fisica Nucleare (INFN), Sezione di Torino, Via Pietro Giuria, 1 10125 Torino, Italy}
\affil[14]{Istituto Officina dei Materiali CNR, Via A. Pascoli snc, 06123 Perugia, Italy}
\affil[15]{Dipartimento di Fisica Scienze Biomediche sperimentali e Cliniche “Mario Serio”, Viale Morgagni 50, 50135 Firenze, Italy}
\affil[16]{Istituto Struttura della Materia CNR, Via Fosso del Cavaliere 100, Roma, Italy}
\affil[*]{Corresponding authors: Leonello Servoli, leonello.servoli@pg.infn.it, Catia Grimani, catia.grimani@uniurb.it}


%

\abstract {The characteristics of a hydrogenated amorphous silicon (a-Si:H) detector are presented here for monitoring in space solar flares and the evolution of strong to extreme energetic proton events. The importance and the feasibility to extend the proton measurements up to hundreds of MeV is evaluated. The a-Si:H presents an excellent radiation hardness and finds application in harsh radiation environments for medical purposes, for particle beam characterization and, as we propose here, for space weather science applications. The critical flux detection limits 
for  X rays, electrons and protons are 
discussed.}




\maketitle

\section{Introduction}\label{sec1}

Space Weather predictions and energetic particle flux monitoring along the orbits of manned and unmanned missions are of paramount importance to preserve astronauts' health, instrument performance in space, and to limit the damages to Earth infrastructures \cite{bertone13,apj1,apj2,apj3,a&aub,cid,taioli23}. The flux of galactic cosmic rays (GCRs) with energies above tens of MeV at 1 au is of the order of 0.1 particles cm$^{-2}$ s$^{-1}$ \cite{gaisser}. The GCR integral  flux above 70 MeV/n has been observed to vary by a factor of four near Earth  during the last three solar cycles  \cite{a&aub}. The overall bulk of observed energetic particles may increase by several orders of magnitude for a few days during gradual solar energetic particle (SEP) events. This increase may occur in less than half an hour in case the solar sources of the events are magnetically well-connected to the point of observations \cite{griele}. Large disturbances of the Van Allen belts associated with the arrival at the Earth of the interplanetary counterparts of coronal mass ejections (ICMEs) and shocks (this last ones also at the origin of SEP acceleration) increase the dose to which are exposed those populations living at high altitude in Bolivia and Argentina in the region of the South Atlantic Anomaly (see \cite{vernetto22} for a work at other latitudes). Some aspects of the physics of ICMEs and associated SEP events and their impact on astronauts and Earth populations are still poorly investigated, despite the NASA and ESA Parker Solar Probe \citep{psp} and Solar Orbiter \citep{a&asolomission, a&asoloscience} missions, among  others, are presently in space to study how the Sun creates and influences the heliosphere. Spatial agencies and scientific communities aim at coordinating  multi-spacecraft continuous monitoring and near-Earth environment observations to understand the role of the interplanetary particle scattering and transport effects to estimate the impact of individual SEP events in the point of observations. Recently, the widespread events  dated  November 29, 2020 \cite{2911} and October 28, 2021 \cite{papa22} were observed with Parker Solar Probe, Solar Orbiter, Stereo \cite{stereo}, Bepi Colombo \cite{bepi}, GOES/SEISS \cite{seiss} and CSES-01/HEPD \cite{martucci23} . These missions returned very different observations in various energy intervals due to different spacecraft orbits. Particle acceleration was observed up to 40 MeV during the November 29, 2020 event while the first ground level enhancement (GLE) of the solar cycle 25 \cite{martucci23} was associated with the October 28, 2021 event. Consequently, the November 29, 2020  event was scarcely interesting from the point of view of impact on manned and unmanned missions, while this was not the case for the October 28, 2021 event.
Multipoint measurements with low-cost, low-weight, low-power-consumption detectors for nanosatellites or long-lived missions in deep space provide precious clues on SEP dynamics, particle acceleration  and, possibly, particle pitch angle distribution. We recall that the SEP pitch angle is defined as the angle between the particle velocity and the nominal interplanetary magnetic field vector direction.  The majority  of solar particle detectors presently hosted onboard space missions aiming at  studying various aspects of Solar Physics allow for particle differential flux measurements up to 100-200 MeV.   Since the typical amount of matter shielding instruments hosted on board space missions deep in the spacecraft is of the order of 10-15 g cm$^{-2}$ (approximately equivalent to the stopping power of astronaut suits) \cite{grim15,a&aub,bridging}, GCR and SEP hadrons with energies below 100 MeV/n play a minor role in the deep charging of spacecraft and dose absorbed by astronauts \cite{bertone13}. For completeness of discussion, we recall that some instruments over the years have covered a larger energy range. Between 2006 and 2016, the magnetic spectrometer PAMELA experiment  gathered  proton and helium differential flux data in a semi-polar elliptical orbit around the Earth above 70 MeV/n during several SEP events \cite{bruno18}, in particular monitored the evolution of the December 13 and December 14, 2006 events \cite{pamFD}. Moreover, some near-Earth instruments such as HEPD on board CSES-01 \cite{bartocci20}, SEISS on GOES and AMS-02 \cite{aguilar18,faldi23} on board the space station  presently allow for the measurement of differential fluxes of solar and galactic protons up to 250 MeV, up to 500 MeV and above 450 MeV n$^{-1}$, respectively. Also
EPHIN on board SOHO \cite{kuhl15,kuhl17} is gathering proton data up to 700 MeV at the first Lagrange point after the mission launch in 1995. The proton differential fluxes measured with EPHIN presents uncertainties  well above 30\% at hundreds of MeV.
 Basically all these experiments are gathering data  at or near 1 au. The comparison of widespread proton observation timeseries in different energy bins carried out with  instruments placed on board distant spacecraft provide precious qualitative information about the propagation of particles. Unfortunately, normalization problems prevent us to infer from these measurements precise differential fluxes. The Solar Orbiter EPD/HET instrument \cite{a&aepd,johan21,sishet,wimmer21} gathers data between 0.28 and 1 au. Despite the HET Collaboration has made efforts to extend the energy range of observations up to 1 GeV,  after more than three years since mission launch, data publicly available on the ESA Solar Orbiter archive\footnote{\url{https://soar.esac.esa.int/soar/}} are limited to 100 MeV energy. The GCR proton data above 100 MeV, kindly provided by the HET Collaboration for September 2020 (private communication), appear higher by approximately a factor of two with respect to those of the AMS-02  experiment \cite{ams_periodicities} gathered above 450 MeV during very similar conditions of solar modulation (October 2019).  An instrument providing  proton flux monitoring up to hundreds of MeV with uncertainties smaller than 30\% if placed on board different spacecraft would improve the present observational scenario.

In this work we  discuss the possibility to build  a new instrument for solar activity and SEP observations up to hundreds of MeV. In particular, we aim at developing a detector in the framework of the INFN HASPIDE (Hydrogenated Amorphous Silicon PIxel DEtectors for ionizing radiation) project, with the sensitive part based on hydrogenated amorphous silicon (a-Si:H). 
The sensing devices are fabricated via the standard PECVD (Plasma Enhanced Chemical Vapor Deposition), with thickness ranging from hundreds of nanometers to tens of micrometers. The substrate resists to total ionizing doses up to 100 Mrad \cite{1999Srour}, displacement damage up to a fluence of 10$^{16}$ 1 MeV neutron equivalent cm$^{-2}$ \cite{DisplacementD2022} and combined damage up to a fluence of 10$^{16}$ 24 GeV protons cm$^{-2}$ \cite{2006Wyrsch}. 
An instrument based on a-Si:H would improve its response in current for extreme SEP events while silicon detectors reduce their performance if exposed to very high doses \cite{sato20}. Furthermore, the possibility of low-temperature deposition during device fabrication, allows us to use  plastic materials like polymmide as substrate. The resulting device will be mechanically flexible, very thin and light-weighted, easily adaptable for curved configurations like cylinders, assuring a large coverage in solid angle.
\par
Several a-Si:H devices have been produced and exposed to different radiation sources: 3 MeV protons,  6 MeV electrons, 6 MV clinical X-ray beams, laboratory X-ray tubes and synchrotron radiation. Preliminary results show a linear detector response in current versus dose-rate, within an uncertainty of 1-2$\%$. Besides, these devices can be operated with electric fields of 1.5-4.0 V/$\mu$m or no bias at all. This latter characteristic is a very important feature for space applications, where power consumption is an important limitation, and for dosimetry in medical physics. These devices, in particular, could be used to equip  missions with low-cost piggy-back modules for the monitoring of SEP events from multiple  observational points  by properly combining active and passive layers of material. 

In Section \ref{secsep} we report the characteristics of SEP events and the present observational scenario. In Section \ref{section:3} the occurrence and evolution of SEP events is presented. In Section \ref{a-si_detect} a-Si:H is presented as detection material. In Section \ref{a-si_test} the results of preliminary test beams with a-Si:H sensors are discussed. In Section \ref{estimate_detect} photon, electron and proton detection limits for a-Si:H devices are estimated on the basis of Monte Carlo simulations. Finally, in Section \ref{demonstrator} the SEP measurement strategy with a-Si:H devices is illustrated.

\section{Solar energetic particles}\label{secsep}
Impulsive, short-term (hours) and gradual, long-duration (days)  increases of solar particles 
are associated with solar flares and CMEs, respectively. Impulsive events are due to magnetic reconnection on open field lines in solar jets \cite{reames21}. CMEs are believed to be generated by sudden disruptions of the Sun’s magnetic field. 
Statistical studies have shown that CMEs with plane of the sky speeds $>$ 1000 km s$^{-1}$ and angular width $>$ 120$^\circ$  accelerate protons above 20 MeV in more than 90\% of the cases, however only a few \% of CMEs have speeds in this range  \cite{lario20,alshehhi21}.
The maximum energy of particles accelerated in impulsive events is about 50 MeV \cite{mazur}.
Pure impulsive or gradual events are rare.
 Often flares and CMEs originate from the same active region of the Sun.
 The evolution of gradual SEP events is characterized by  particle
flux rises over a minimum period of  half-an-hour for events magnetically well connected to the point of measurement  and a slow decay of a few days.
A strong increase after a first decay phase observed during the most intense events 
is due to superposing CMEs and shocks. The majority of solar proton events
occurs during years 5-8 of the solar cycle \cite{shea01}. It has
to be pointed out that, in most cases, SEPs present energies smaller than 10
GeV even if observations of particles with energies as high as 50 GeV and
more have been reported in the literature. In particular,  integral proton
intensities of about 10$^{-5}$ (cm$^{2}$ s sr)$^{-1}$ at energies $>$ 500 GeV have been measured
for the September 29, 1989, June 15, 1991, and October 12, 1981 GLEs  \cite{karpov}. 


The Archimedean spiral pattern of the interplanetary magnetic field generates an asymmetry in the intensity-time profiles of SEP events from eastern and
western longitudes on the Sun. In particular, events originating in the western
hemisphere of the Sun are more likely observed at the
Earth  with respect to those from the eastern hemisphere. The
properties of SEP profiles associated with gradual events are explained in terms of direct magnetic connection
between the shock driven by the ICMEs and the detecting spacecraft. Here and in the following for onset  of SEP events we mean the time when the integral flux of particles above several tens of MeV overcomes the background of galactic cosmic rays beyond statistical fluctuations and short-term variations. The peak is when the integral flux is maximum in the same energy range. The time of the onset corresponds to the period when particles are observed at the highest energies and it is believed to correspond to the time at which the shock intercepts the magnetic field lines to
the spacecraft, plus the particle propagation time to the spacecraft. The time of the
maximum integral flux is believed to be an indication of the time at which the spacecraft is connected
to the part of the shock which accelerates the particles more efficiently \cite{dalla03}.
Particle acceleration at parallel and perpendicular shocks occurs when the upstream magnetic field is almost parallel or orthogonal to the shock normal, respectively.  Parallel shocks are considered to be the main SEP accelerators (see \cite{joyce21} and references therein). Acceleration at perpendicular shocks, in comparison to parallel shocks, has been discussed in \cite{zank06}, for instance. A particle velocity dispersion is observed during the first phase of the events when particles arrive in inverse order to their velocity (an electron flux increase is observed before a proton flux increase), while identical time profiles are observed during the final part of the event. 
During gradual, magnetically well connected events,  particles accelerated beyond 1 GeV are detected within tens of minutes after the visual recognition of the event on the Sun. Low-energy particles appear in increasing numbers during the evolution of the event while the high energy particles
fade away.
For many years all our knowledge on solar phenomena was acquired from
the viewpoint of the Earth that moves between $\pm$7.25$^{\circ}$ in heliolatitude.
The comparison of Ulysses  \cite{ulysses1,ulysses2}, STEREO A and STEREO B \cite{lario13}, Helios 1 and 2 \cite{lario06,qin15} and more recently Parker Solar Probe, Solar Orbiter and Bepi Colombo 
observations, along with near-Earth data have provided precious clues about  SEP propagation as a function of
solar latitude, longitude and distance from the Sun. 

Due to the paucity  of instruments in space devoted to the detection of particles with energies larger than 100-200 MeV and the importance to measure SEP fluxes from MeV to the highest possible energies, we aim at building a robust detector to be hosted on long-duration missions in deep space or in harsh environments such as the Van Allen belts for the monitoring of strong-to-extreme SEP events and for the study of the effects of the Van Allen belt discharging due to ICME transit \cite{zhao19,adriani2015}.

\section{Solar energetic particle event occurrence and forecast \label{section:3}}
The SEP event average occurrence as a function of fluence and solar activity can be estimated according to the Nymmik's model \cite{nym99a,nym99b}. The impact of ICMEs and SEP events on manned and unmanned space missions and Earth infrastructures must be studied on the basis of both particle fluence and flux. As a matter of fact, events with the same fluence can be associated with particle fluxes differently populated at high energies. 

\subsection{X-ray and electron emission from the Sun for SEP event forecast} 

\begin{figure}[h]%
\centering
\includegraphics[width=0.7\textwidth]{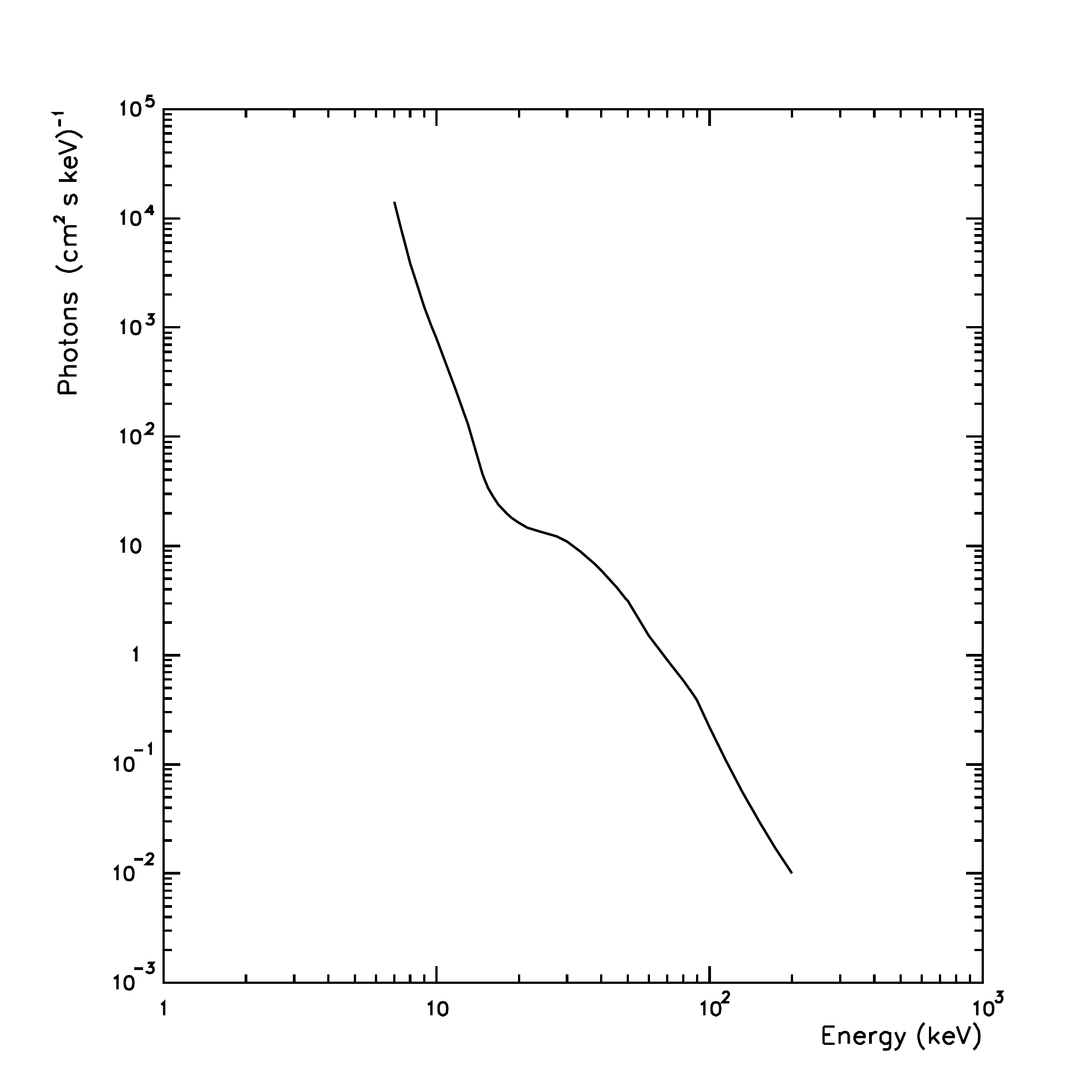}
\caption{
X-ray  spectrum of a class M flare observed with the RHESSI satellite on November 9, 2002 at 13:14:16 UT \cite{benz17}.}
\label{fig:hardsoft}
\end{figure}

Observations of strong emissions of soft X rays and near-relativistic electrons from the Sun can be used to forecast SEP events \cite{rigo16,nunez18}.
In Fig. \ref{fig:hardsoft} we have reported the measurement  of the X-ray spectrum carried out by the RHESSI experiment \cite{grigis04,benz17} during a class M flare dated November 9, 2002. This observation is used in the following as a benchmark for the performance of the instrument we propose. 

\begin{figure}[h]%
\centering
\includegraphics[width=0.7\textwidth]{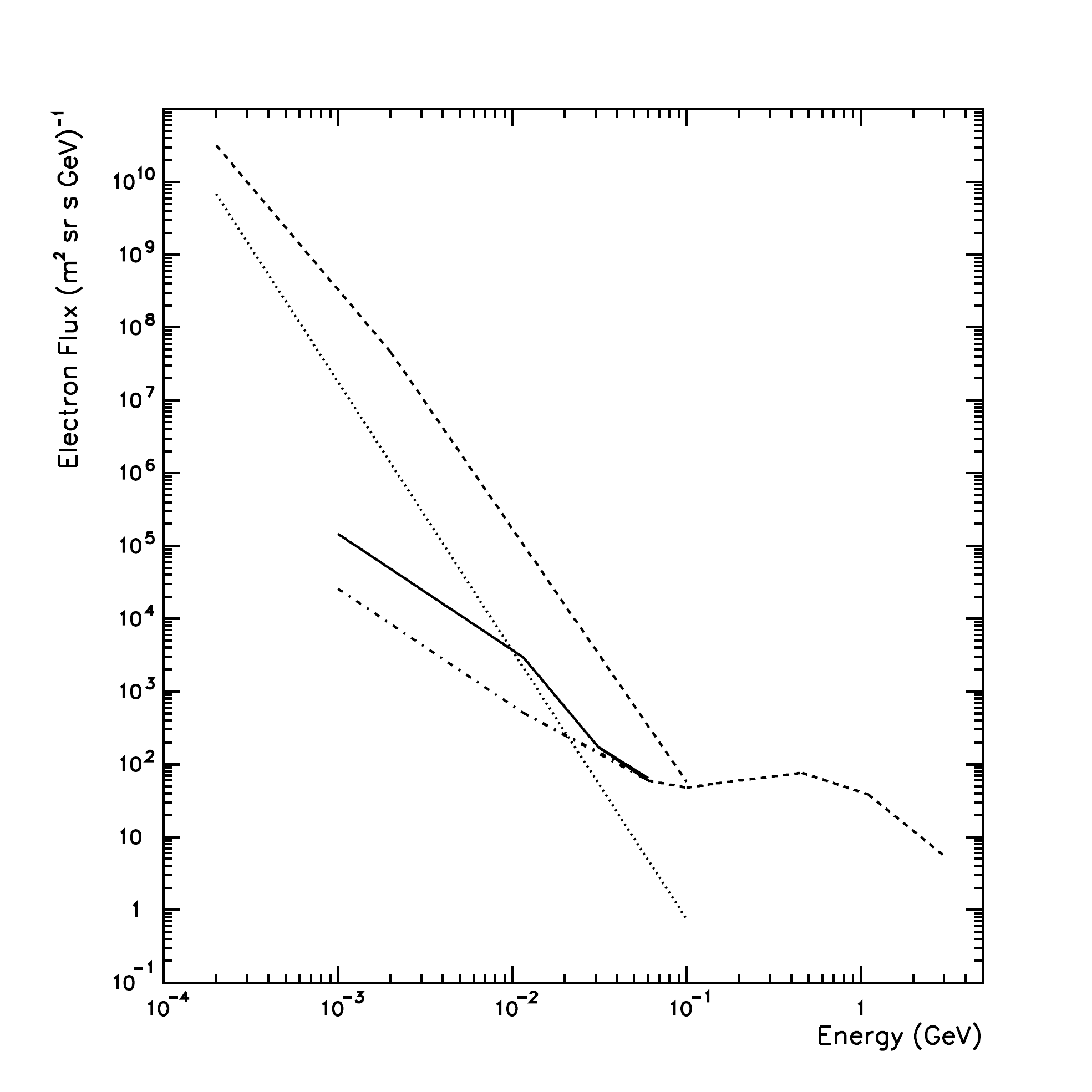}
\caption{Solar, interplanetary and galactic electron observations near Earth.  The top dashed line represents the electron flux of solar origin observed during the impulsive solar event dated November 7, 1973. The dotted line indicates the electron flux observed during the gradual event dated September 3, 1973. The continuous and dot-dashed lines are the  minimum and maximum components of electrons of Jovian origin while above about 30 MeV the maximum component of electrons of galactic origin is reported during the negative polarity period of the global solar magnetic field. This figure was adapted from Fig. 7 in \cite{griele}. }
\label{fig:ele_sol}
\end{figure}

Interplanetary electrons consist of several components \cite{griele}. At steady state, below 30 MeV the majority of electrons are accelerated in the Jovian magnetosphere. The dot-dashed and continuous lines in Fig. \ref{fig:ele_sol} represent the minimum and maximum fluxes of electrons of Jovian origin near Earth. The Jovian electron flux change is due to the varying distance between Earth and Jupiter during the Jovian synodic year. Above 30 MeV electrons are mainly galactic. The dashed line above 30 MeV represents the maximum component of galactic electrons near Earth observed at solar minimum during negative polarity periods of the global solar magnetic field.
During SEP events,  relativistic electrons reach the point of observations always before non-relativistic ions. According to Posner \cite{pos07} (see also \cite{2009Grimani}), for events with intensities larger than 10 protons (cm$^2$ sr s)$^{-1}$ above 10 MeV,  relativistic MeV electrons and non-relativistic 50 MeV proton fluxes show  similar time profiles even though at different times due to the higher velocity of the electrons.   Posner points also out that if the proton increase is not observed after 2-3 hours from electron increase, then it will not be observed at all. 
The top solar electron energy spectrum (dashed line) in Fig. \ref{fig:ele_sol}, characterized by two different spectral indices, represents the electron flux of solar origin observed during the impulsive solar event dated September 7, 1973.  The dotted line represents the energy spectrum  of electrons observed during the gradual event dated November 3, 1973.  Only one electron flux spectral index is observed during gradual events resulting from a process of shock acceleration. The dashed and dotted lines were obtained from data interpolation \cite[see][for details]{griele}. 

\subsection{Solar proton energy spectra}
The observed energy spectra  of solar particles
 vary during the evolution of the events, as a consequence of particle acceleration and propagation in the interplanetary medium.  In particular, the majority of SEP energy spectra above tens of MeV show an exponential trend  during the prompt phase and a power-law trend during the late
phases of the majority of events (for a detailed study of the evolution of several SEP events evolution see \cite{grimani2013}).
The comparison of the proton timeseries in different energy bins allow us to study the arrival of particles as a function of their energy. For this kind of study any observation, above the galactic cosmic-ray background, provides precious clues on the particle propagation process and the qualitative comparison of data gathered with different  spacecraft is meaningful. Nevertheless, to estimate  spacecraft deep charging and instrument performance with Monte Carlo simulations, it is necessary to monitor the high-energy differential flux of particles versus time. Moreover, in order to obtain plausible results from the simulations, the uncertainty on the high-energy particle flux, which is the part of the flux that impacts the most in terms of deep charging and dose-rate release, must be kept low. The effort of comparing observations gathered with different instruments, in general, is unsuccessful  due to particle flux normalization issues. To this purpose, in order to estimate the number of photons and particles that the instrument we propose may observe, in the following we privilege the flux of  particles measured by single instruments, if possible. 
In particular we consider the energy spectra of solar protons associated with the evolution of SEP events dated February 23, 1956 \cite{vashe}, December 13, 2006 \cite{adriani16}, and October 28, 2021 \cite{papa22} 
as case studies. The proton spectra parameterizations at the onset and at the peak of the February 23, 1956 event are provided in the original paper \cite{vashe}, while for the December 13, 2006 event the reduced $\chi^2$ has been calculated on the basis of the data reported in the cosmic-ray database of the Italian Space Agency\footnote{\url{https://tools.ssdc.asi.it/CosmicRays/}}. For the event dated October 28, 2021, it is pointed out that the data and uncertainties of the EPD/HET instrument are reported on the Solar Orbiter archive. Within data uncertainties (100\%) the fits present an excellent agreement. The best fits through the data during the different phases of each event were found by considering exponential, power-law and  power-law with an exponential cut-off functions according to \cite{grimani2013}. The results are shown in Table \ref{tab2}.

\subsubsection{The February 23, 1956 SEP event}
The SEP event dated February 23, 1956 was associated with the GLE 5 when  neutron monitor (NM) counting increased  by about 5000\%  \cite{vashe}. 

\begin{figure}[h]
\centering
\includegraphics[width=0.8\textwidth]{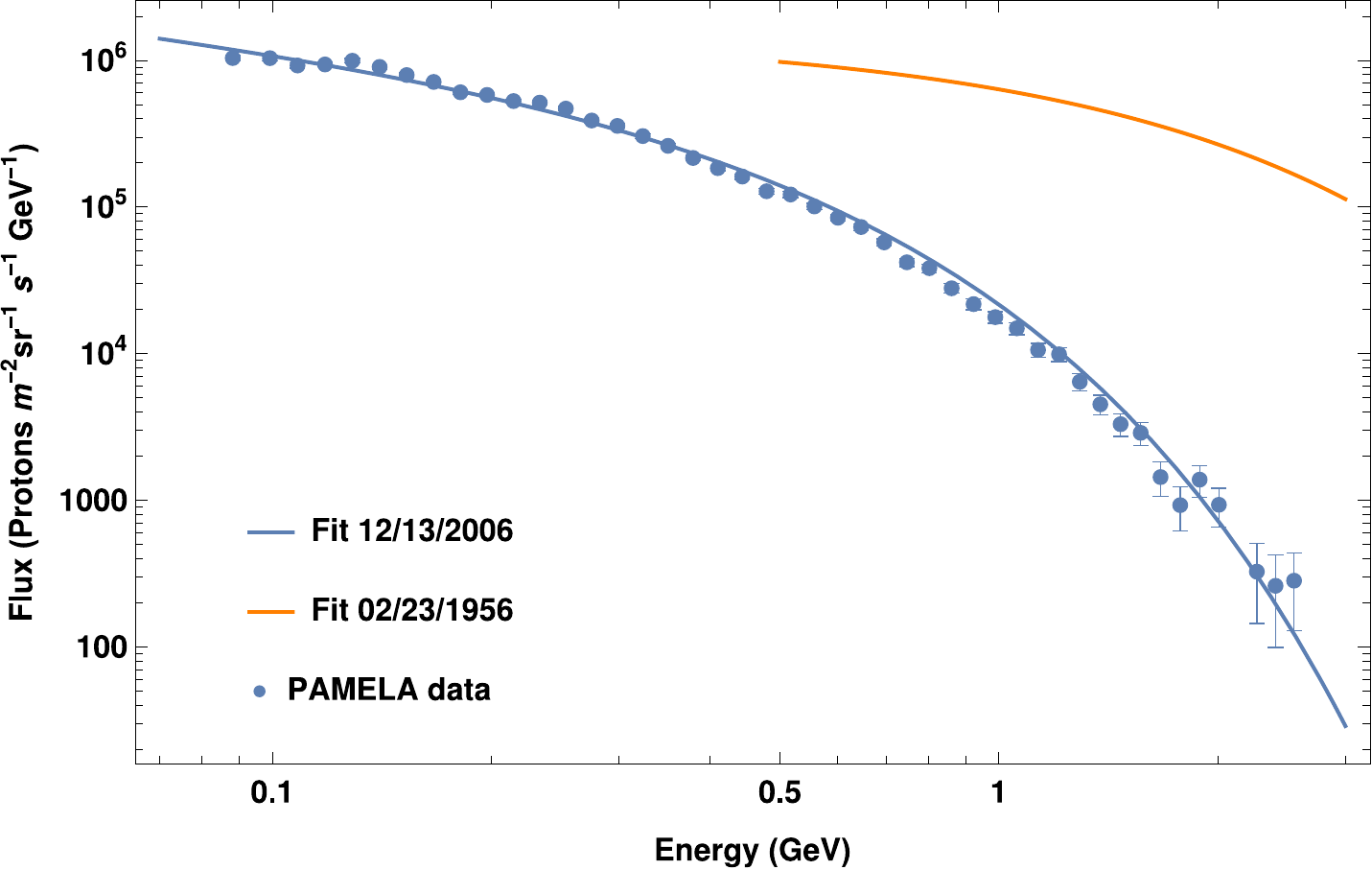}
\caption{Proton flux at the onset of the SEP events dated February 23, 1956 and December 13, 2006. The curves represent the best fits to the data (see Table \ref{tab2}). 
}
\label{fig:sep_onset}
\end{figure}

\begin{figure}[h]
\centering
\includegraphics[width=0.8\textwidth]{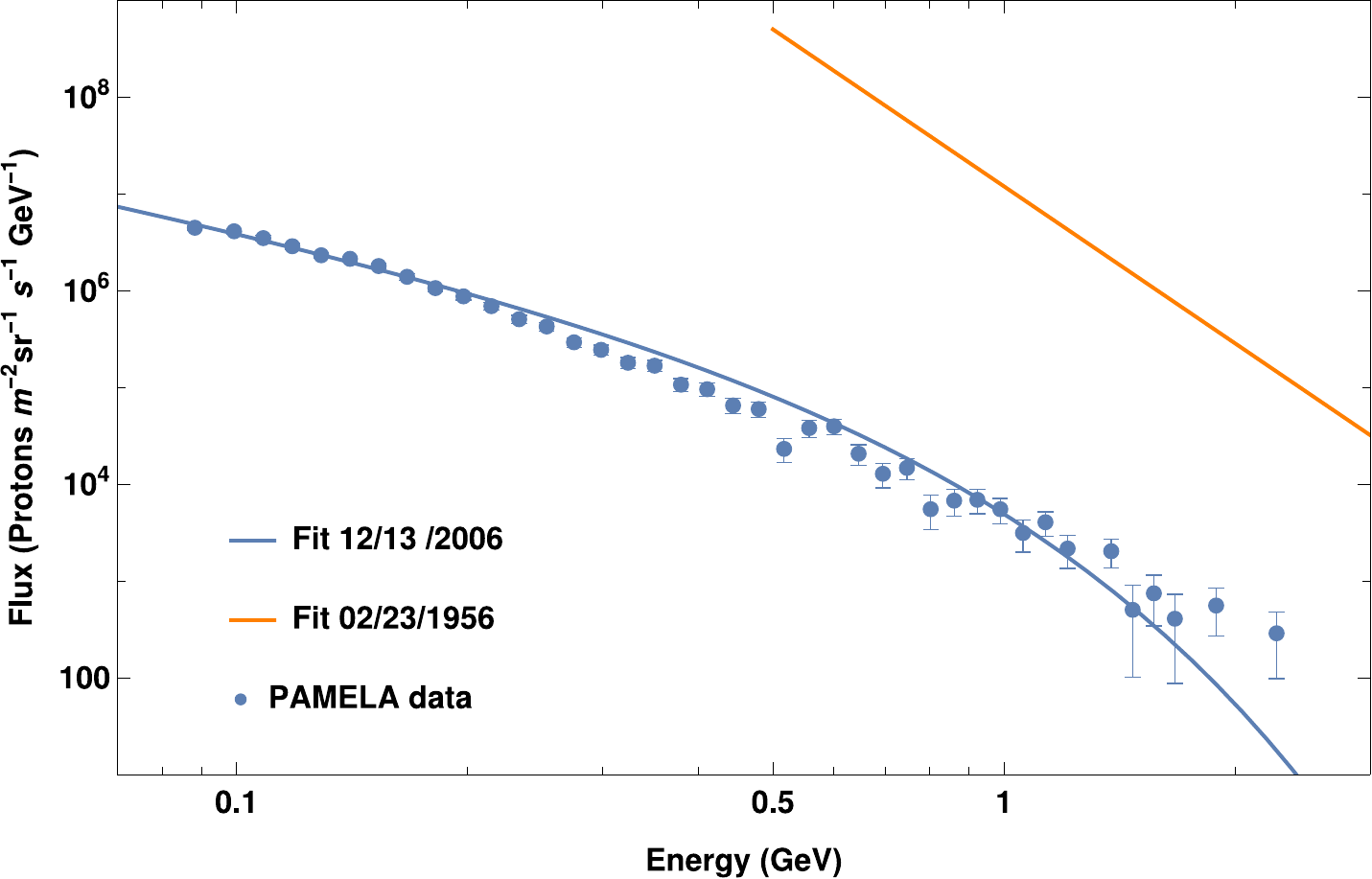}
\caption{Same as Fig. \ref{fig:sep_onset} at the peak of the SEP events.}
\label{fig:sep_peak}
\end{figure}

This is the most intense event observed during the last seventy years characterized by particle acceleration above 2 GeV. 
The event onset was observed with NMs at 04:00 UT while the decay phase was detected at 06:00 UT. The fitting functions at the onset and at the peak of the event appear in Figs. \ref{fig:sep_onset} and \ref{fig:sep_peak} and in  Table \ref{tab2}. 

\begin{sidewaystable}
    \centering
    \caption{Parameterizations of proton energy spectra during the evolution of  SEP events of different intensities (see Figs. \ref{fig:sep_onset}-\ref{fig:onset2810}). $E$ indicates the proton  energy in GeV. The spectra are meant in protons (m$^2$ sr s GeV)$^{-1}$. Reduced $\chi^2$ ($\chi^2_{red}$) and number of degrees of freedom (ndof) have been reported as an estimate of  the goodness of the  data fitting except for the February 23, 1956 event for which we have adopted the parameterization of the proton fluxes reported in the original paper and where no data points were shown \cite{vashe}. Measurement instruments are indicated.}
    \label{tab2}
    \begin{tabular}{lcccc}
    \toprule
    &&&$\chi_{red}^{2}$&ndof\\
    \midrule
    \multicolumn{5}{c}{February 23, 1956 (inferred at the top of the atmosphere from NM data)}\\
    \midrule
    $Onset$ 4:00 UT & $1.5\times10^6 \exp(-E/1.16)$ && -& -\\
    $Peak$ 4:30 UT & $1.2\times10^7 E^{-5.39}$&& -& -\\
    \midrule
    \multicolumn{5}{c}{December 13, 2006 (Pamela satellite experiment)}\\
    \midrule
    $Onset$ 3:18-3:49 UT & $4.47\times10^5 \exp(-E/0.33) E^{-0.51}$ &&1.26&41\\
    $Peak$ 4:33-4:59 UT & $1.54\times10^5 \exp(-E/0.29) E^{-1.54}$&&1.72&37\\
    \midrule
    \multicolumn{5}{c}{October 28, 2021 (Solar Orbiter/HET instrument)} \\
    \midrule
    \multirow{4}{*}{$Onset$ 15:35-16:35 UT}& $1.32\times10^9 E^{2.81}$& $0.020\leq E \leq 0.042 $&0.0050&7 \\
    &$2.12 \times10^6 E^{0.78}$&  $0.042\leq E \leq 0.080 $ & 0.0084& 6\\
    &$7.50\times10^6 \exp(-E/0.135) E^{1.06} $& $0.080\leq E \leq 0.90 $& -& -\\
    &$5.73\times10^3 E^{-4.58}$& $0.90\leq E  \leq 2.41$& 1.12& 7\\
    \\
    \multirow{4}{*}{$Peak\; (NM)$ 17:30-18:20 UT}&  $9.66\times10^6 E^{-0.0875}$ & $0.020\leq E \leq0.040$&  0.0012& 7\\
         &$6.45\times10^5 E^{-0.93}$& $0.040\leq E \leq 0.073$& 0.0044& 5\\
         &$1.03\times10^6 \exp\left( -\frac{E}{0.157} \right) E^{-0.94} $& $0.073 \leq E \leq 0.480$& -& -\\
         &$2.55\times10^3 E^{-4.72} $& $0.480\leq E \leq 2.250$& 2.0344& 12\\
    \\
    \multirow{3}{*}{$Peak\; (space)$ 20:35-22:35 UT} & $3.11\times10^5 E^{-1.30}$& $0.020\leq E \leq 0.036$& 0.0093& 5\\
    &$4.18\times10^4 E^{-1.90} $&$0.036 \leq E \leq 0.072$& 0.0021& 6\\
    &$1.58\times10^4 E^{-2.27} $&$0.072 \leq E \leq 0.580$& 0.0007& 1\\
    \botrule
    \end{tabular}
\end{sidewaystable}

\subsubsection{The December 13, 2006 SEP event}
The  measurements of  SEP events  carried out  on December 13 and December 14, 2006 with the PAMELA low-Earth orbit satellite experiment covered the range of energy 70 MeV-2 GeV \cite{pamFD} with an unprecedented accuracy. In Figs. \ref{fig:sep_onset} and \ref{fig:sep_peak} we have  reported the onset and peak proton fluxes (light blue dots) associated with the strongest of the two events dated December 13, 2006. 
The fitting functions of the proton fluxes measured during this event are also reported in Table \ref{tab2}. The reduced $\chi^2$ ($\chi^{2}_{red}$) and the number of degrees of freedom (ndof) for each function appear in the same Table.
The event dated December 14 was very weak and it is disregarded here.

\begin{figure}
\centering
\includegraphics[width=0.7\textwidth]{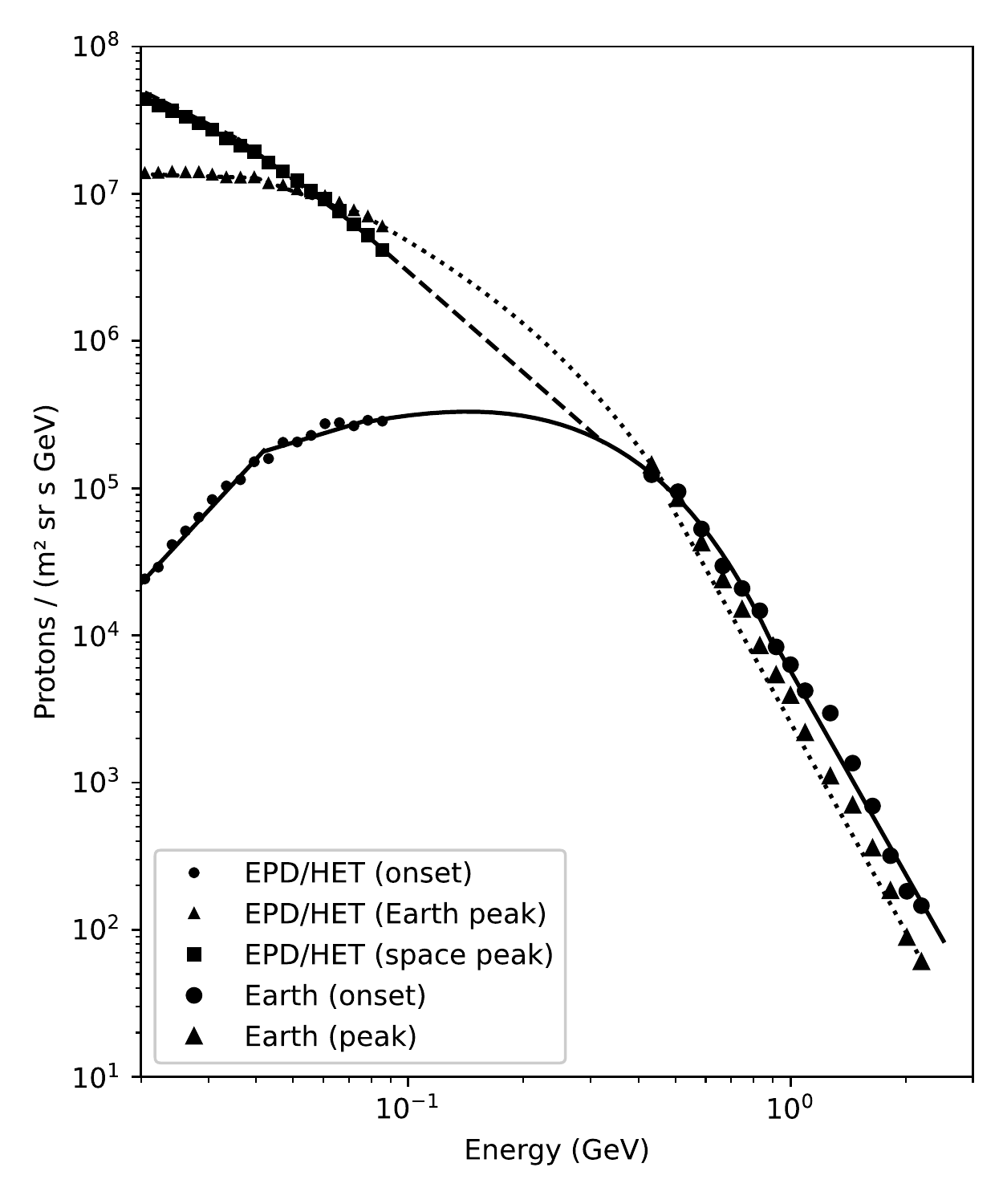}
\caption{Solar energetic proton fluxes observed on October 28, 2021. The small solid dots indicate the observations gathered in space by the EPD/HET experiment at the onset of the event. The large dots inferred at the top of the atmosphere from NM observations above 450 MeV  represent Earth observations for the same phase of the event. The small triangles are the HET measurements in space at the peak of Earth observations (large triangles). The solid squares correspond to the peak of the HET measurements in space. The lines through the data represent the best-line-fits in different energy intervals for each phase of the event (see Table \ref{tab2}).}
\label{fig:onset2810}
\end{figure}

\subsubsection{The October 28, 2021 SEP event}\label{3.proton}
On October 28, 2021 a SEP event was detected by several spacecraft at different distances from the Sun, by near-Earth space missions and with  NMs, after using cosmic-ray nuclear transport in the atmosphere  \cite{papa22}. The atmosphere does not stop secondary particles only if primary cosmic rays have energies larger than 500 MeV.  Protons accelerated above 2 GeV were associated with this event. 
In Fig. \ref{fig:onset2810} we have shown the proton flux measured during the evolution of this event. Data below 100 MeV were gathered by the EPD/HET particle detector hosted on board the Solar Orbiter spacecraft that was almost lined up with Earth. HET data at the onset (small dots), at the peak on Earth  (small triangles)  and at the peak in space (small squares) are compared to data estimated with NM observations  at the onset (large dots) and at the   peak at the top of the Earth atmosphere (large triangles). No GLE was observed at the time of the peak observed in space (small squares).  In other words, the peak of the event observed in space was  associated with particles with energies smaller than 500 MeV. These observations reveal the importance of monitoring SEP events in space up to at least hundreds of MeV  since this is the energy range of maximum interest to Space Weather. We have best fitted the data with broken lines where a unique function was not reproducing well the trend of the observations. The  parameterizations of the proton energy spectra are reported in Table \ref{tab2} with the reduced $\chi^2$ and number of degrees of freedom. Reasonable interpolation functions have been used to fill the gaps of missing data.
\begin{figure}
\centering
\includegraphics[width=0.9\textwidth]{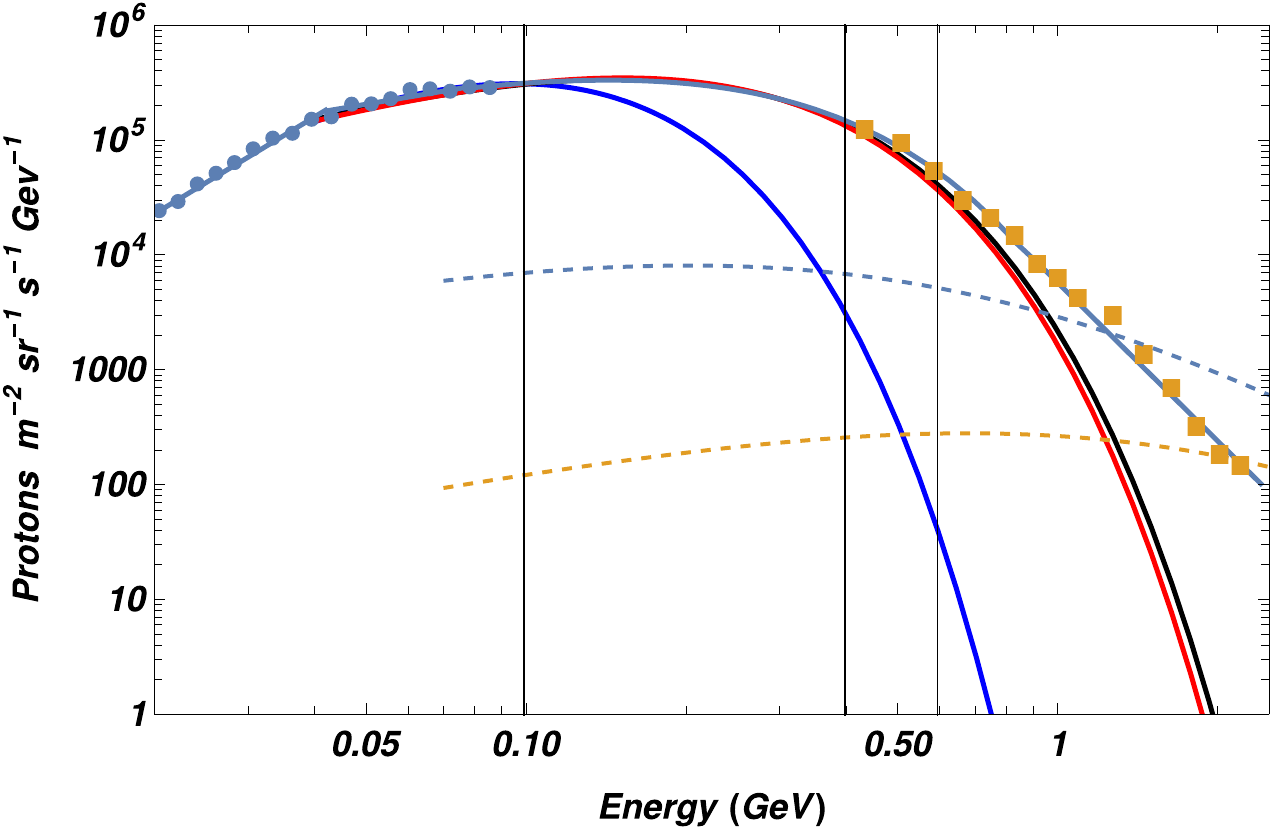}
\caption{EPD/HET data at the onset of the October 28, 2021 event (solid light blue dots). Data inferred from NM observations during the same phase of the event (solid orange squares). The overall flux has been parameterized with broken lines through the data (light blue curves). The thick blue line represents the best fit to the EPD/HET data above 50 MeV but below 100 MeV. If data from 50 MeV to 400 or 600 MeV are considered for the fit the red and black curves are obtained. The dashed curves represent the galactic cosmic-ray background at solar minimum (light blue) and solar maximum (yellow).}
\label{fig:fit_onset_n}
\end{figure}

In Figs. \ref{fig:fit_onset_n} we have reported the October 28, 2021 SEP data at the onset of the event.
The HET data have been parameterized between 50 MeV and 100 MeV (blue line), between 50 MeV and 400 MeV (red line) and between 50 MeV and 600 MeV (black line). Above 100 MeV the interpolation values  obtained by considering the high-energy NM measurements extrapolated at the top of the atmosphere (light blue line) were used as input data. It is possible to notice that, in this last case, a quite good agreement with data is found up to 1 GeV when galactic cosmic rays overcome those of solar origin at solar minimum. 

\begin{figure}
\centering
\includegraphics[width=0.9\textwidth]{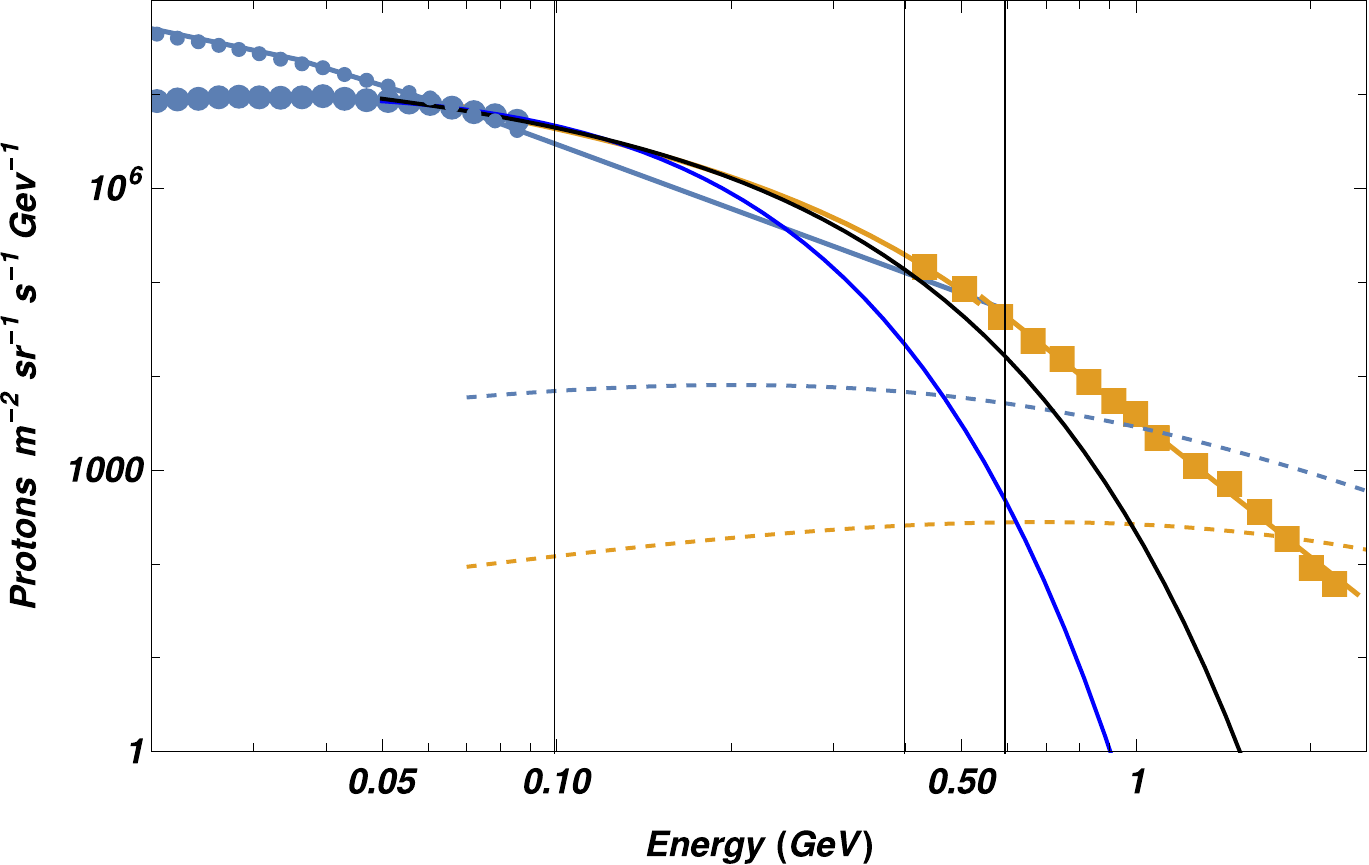}
\caption{EPD/HET data gathered during the October 28, 2021 SEP event at the peak of the event observed in space (small light blue dots) and at the peak on Earth obtained from NM data extrapolated to the top of the atmosphere (large light blue dots and orange squares). The light blue curve represents the best fit to the data corresponding to the peak of the event in space. The fit cannot be reasonably extended above 600 MeV without other measurements. The thick blue line represents the parameterization of EPD/HET data (large dots) gathered in space. Data above 50 MeV and below 400 MeV or 600 MeV allow for the same parameterization (black line). These parameterizations are meant to be compared with the  orange curve that represents the best fit to the space observations and data gathered with NMs during the time the peak of the event was observed on Earth. The dashed curves have the same meaning as in Fig. \ref{fig:fit_onset_n}.}
\label{fig:fit_peak_n}
\end{figure}

In Fig. \ref{fig:fit_peak_n} we report the observations gathered by HET during the peak of the event in space (small light blue dots) and on Earth with NM observations extrapolated at the top of the atmosphere (large light blue dots) during the same event. The proton peak in space is parameterized up to 600 MeV.  Also in this case, by including in the parameterization  the data up to 400 or 600 MeV, a better agreement is found (Fig. \ref{fig:fit_peak_n}). 
In conclusion, in order to carry out reliable estimates of SEP fluxes at GeV energies, low uncertainty differential flux measurements between 50 MeV and 400-600 MeV are needed. This explains and motivates our maximum effort to design and build a new instrument to measure SEP fluxes in space up to hundreds of MeV.

\section{a-Si:H as detection material}\label{a-si_detect}

Hydrogenated amorphous silicon has been used since many years in the fabrication of devices related to optoelectronics, such as solar cells, thin-film transistors and other applications \cite{aSIh}. Several methods have been proposed for the preparation of device-grade a-Si:H but the PECVD method is the most widely used due to its capability to consistently prepare uniform, high-quality materials on
a large-area substrate \cite{2017Matsuda}. 
The amorphous silicon is a material highly resistant to ionizing radiation damage due to its intrinsic disordered nature, but to obtain a detector grade device is necessary to reduce the number of dangling bonds inside the material. This is done by introducing H into the material to passivate dangling bonds hence reducing defects and recombination centers. The minimum amount of H necessary to passivate most of the dangling bonds is about 1\% atomic. The increase of H  content enlarges the bandgap hence reducing the background current of the device, and $\sim$ 14\% is the typical value to obtain a detector grade device \cite{wyrsch16}. The bandgap depends also on the deposition conditions such as the temperature. Typical deposition temperature should range between 100~$^{\circ}$C (to reduce the number of defects in the material) and 350~$^{\circ}$C (to avoid desorption of H), and normally occurs around 200~$^{\circ}$C \cite{2017Matsuda}. This interval of processing temperature allows for the deposition of layers or devices on a variety of substrates, among which thin layers of plastic materials like polymmide \cite{soder2008,2023Menichellitubino}. The capability of depositing small area devices over a flexible and thin substrate opens the possibility to several applications in different fields among which: instrumented flange at the vacuum/air separation of charged particle accelerators and transmission detectors for dosimetry of real dose delivered to patients by clinical beams in radiotherapy, by considering curved or bent  small thickness devices of almost any geometrical shape suitable also for space applications. 
\par
\begin{figure}[h]%
\centering
\includegraphics[width=0.9\textwidth]{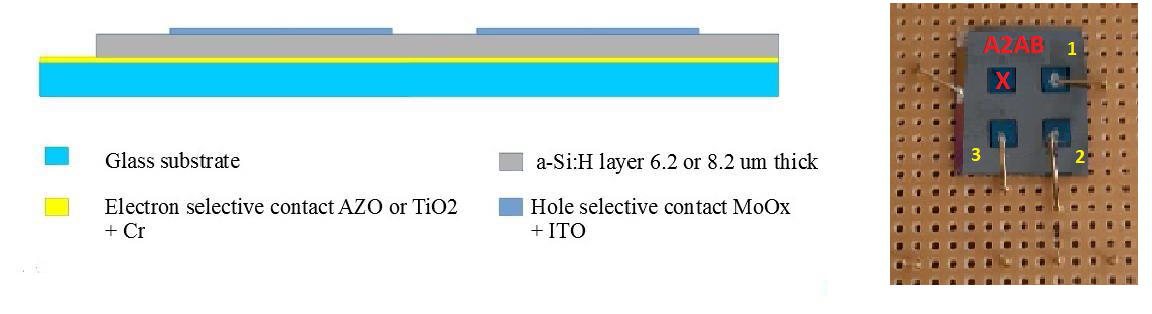}
\caption{Prototype devices A2AB: 4 diodes with charge selective contacts over a glass substrate; 4x4 mm$^2$ area, 8.2 $\mu$m thickness; (left) single diode scheme;  (right) picture of the device mounted on electronic board for testing. Diode marked with X is not working.}
\label{fig:device_A2AB}
\end{figure}

Within the research program of the HASPIDE project several non-optimized devices have been fabricated and tested both with and without ionizing radiation beams.
Fig. \ref{fig:device_A2AB} shows 4 diodes of 4x4 mm$^{2}$ area and 8.2 $\mu$m thickness deposited within the same fabrication batch. 
Two types of contacts have been used: standard p-i-n diodes, with the two contact layers, respectively p-doped (with B atoms) and n-doped (with P atoms), separated by the a-Si:H intrinsic layer and Charge Selective Contacts (CSC) where for the two contact layers are used metallic oxides, MoOx for hole contacts and TiO2 or ZnO:Al (AZO) as electron contacts, to produce different mobility values for electrons and holes \cite{2022Menichelli}. The devices are biased and readout by a precision source meter to record the current with a measurement frequency of $\sim$ 1 Hz.
\par
An important property for the foreseen medical and space applications is the capability of devices to work at $\sim$ 0 V bias. Both contact options allow in principle for this feature that should be thoroughly investigated.

\section{Tests of a-Si:H devices with photon, electron and proton beams}\label{a-si_test}

Several devices, both p-i-n diode and CSC have been characterized for noise, leakage current and with ionizing radiation beams at several facilities:  X-ray photons (tube voltage up to 50 kV), 6 MV clinical photon beams, 6 MeV  electron beams and 3 MeV proton beams \cite{2022Menichelli,2021Menichelli}.\footnote{Facilities used for beam tests: for photons below 50 keV see \cite{2023Menichellitubino}.\\ For photons in the MeV range: ANSTO Melbourne, \url{https://www.ansto.gov.au/facilities/australian-synchrotron}.\\ For protons: CEDAD Brindisi, \url{http://www.cedad.unisalento.it/}.\\ For electrons: \url{https://www.sbsc.unifi.it/.}} 


The procedure for measuring the current signals from a device under test is: (i) start the data acquisition with no irradiation to measure the leakage current, (ii) irradiate the sensor (with X-ray photons, for instance), and finally (iii) measure again the background noise, after irradiation.



\begin{figure}[h]%
\centering
\includegraphics[width=0.7\textwidth]{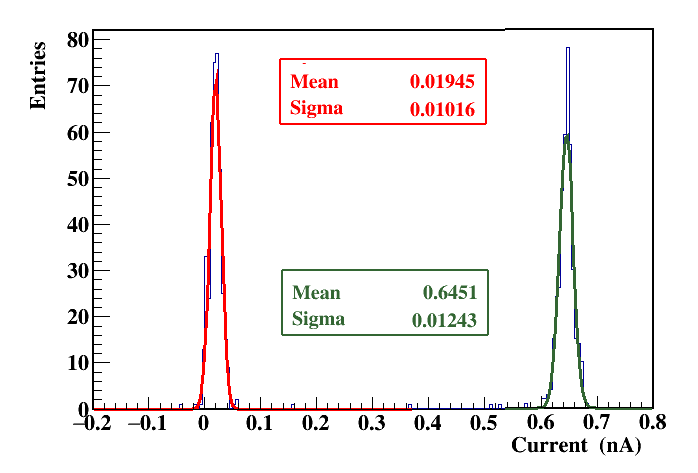}
\caption{Noise determination for both background current (red peak) and signal (green peak) during irradiation. The noise is represented by the width of the gaussian peaks.}
\label{fig:Noise}
\end{figure}


The uncertainty on current measurements (packaging included), is determined from the distribution of the measurements when the beam is off (red peak in Fig.\ref{fig:Noise}) and on (green peak in Fig.\ref{fig:Noise}). The distribution for the device under test shows a background current of the order of 20 pA with a noise of 10 pA including the contribution of the non-optimized setup, while in the presence of the X-ray beam the current rises up to hundreds of pA while the noise increases marginally (12 pA) including both the sensor response uncertainty and the beam stability. The current signal, due to the exposure of a device to radiation, is obtained by subtracting the background current, measured before the beam is switched on, from the measurement carried out after stabilization.

\par
The homogeneity of devices produced in the same production batch follows from the deposition procedure and is $\sim$ 20$\%$ at worst for diodes in Fig. \ref{fig:device_A2AB}, reaching also 5$\%$ for more recent batch of diodes deposited on kapton. The variance for the entire bias range is smaller than 20$\%$, ensuring that the same readout chain would work for all devices after inter-calibration \cite{2023Menichellitubino,large23}.

 For this work, it is important to evaluate the detection limit of the signal, by determining the measurement uncertainty. This  evaluation is carried out by looking at the fluctuations of the background current that is influenced by the setup and the measurement instrumentation. During the testing campaign we have found that the  current noise ranges between 2 and 10 pA, depending mainly on the measurement setup rather than on the sensor characteristics. In the following we will adopt 5 pA as a reasonable average value of the whole testing campaign and as a worst case of  the most  recent optimized results (see Table \ref{table:types}).

\begin{table}[h]
\begin{center}
\caption{Characteristics of devices exposed to different type of ionizing radiation. V2 and V4 identify two different pads mounted on the same socket.}\label{table:types}%
\begin{tabular}{lllllll}
\toprule
Sensor Name & PAD$\_$(V2,V4) & A2AB1 & A3AC2 & UOW429 & \\
\midrule
Contact type    & p-i-n    & CSC  & CSC & p-i-n & \\ 
Area   [mm$^{2}$]    & 0.5x0.5    & 4.0x4.0  & 4.0x4.0 & 5.0x5.0 \\
Thickness  [$\mu$m]  & 10.0   & 6.2  & 8.2 & 10.0  \\
Noise level [pA]   & 2.0   & 8.0  & 5.0 & 2.5  \\
\botrule
\end{tabular}
\end{center}
\end{table}

 If we define the detection limit at 5$\sigma$ for a typical device, this would imply 25 pA of minimum current signal.

\begin{figure}[h]%
\centering
\includegraphics[width=0.7\textwidth]{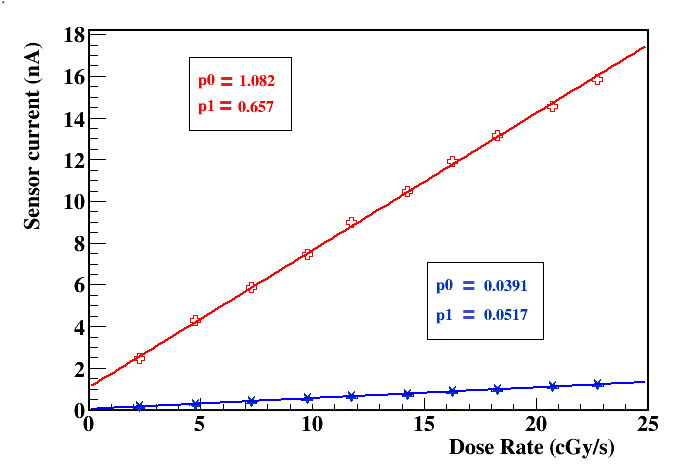}
\caption{Linear correlation between dose rate and signal for A2AB1 CSC device at 0 V bias (blue) and 20 V bias (red). The two parameters p0 and p1 represent the intercept and slope, respectively, of the linear fits to the data shown in the plot as continuous lines.}
\label{fig:Xrayscan}
\end{figure}

Fundamental properties for the present study of a-Si:H meant for  Space Weather application are the device noise, the linearity of the response to the ionizing radiation flux and the device sensitivity, i.e. the current signal associated with a given amount of deposited energy. The linearity has been measured using several devices (pixel, strips, pad) with different contact types (CSC or p-i-n) (see Table \ref{table:types}) over a variety of beam types. 
As an example, 
Fig. \ref{fig:Xrayscan} shows the relation between dose-rate (measured using a certified dosimeter) and signal for a CSC device of 4x4 mm$^{2}$ area, 8.2 $\mu$m thickness for two potential bias values of 0 V (blue) and 20 V (red) and irradiation with an X-ray beam. See also \cite{Menichelli22}. The two bias values represent a reasonable range for thin devices.

The linearity of the response over the dose-rate range is on average within 1.6$\%$ for both biases, with a maximum distance from the fit of 5$\%$ for low dose rate values. The sensitivity varies from 657 pA$\cdot$ cGy$^{-1}$ s$^{-1}$ (20 V bias) to a value of 52 pA$\cdot$ cGy$^{-1}$ s$^{-1}$ (0 V bias), one order of magnitude less. Since the current noise has been assumed of 5 pA, if the measurement threshold is set at 5$\sigma$, the minimum energy deposition that we could measure varies from $\sim$ 0.4 mGy s$^{-1}$ for 20 V bias to $\sim$ 0.5 cGy s$^{-1}$ for 0 V bias operation.
Similar results were obtained for p-i-n devices and are available in \cite{2021Menichelli}.  Table \ref{tab:devices} reports the results obtained with several sensors exposed to different ionizing radiation beams.  

\begin{table}[h]
\begin{center}
\caption{Response of devices exposed to different type of ionizing radiation.}\label{tab:devices}%
\begin{tabular}{lllllll}
\toprule
Ionizing & Sensor  & Bias  & Sensitivity & 5$\sigma$ detection & Linearity \\
Radiation & Name & [V] & [nC cGy$^{-1}$] & value   [cGy] &   [\%]\\
\midrule
$\gamma$ X-ray  & A2AB1   & 0.0 & 0.052  & 0.260 &  $\sim$ 1.6  \\
$\gamma$ 50 kV X-ray  & A2AB1   & 20.0 & 0.657 & 0.040 & $\sim$ 1.6  \\
$\gamma$ 50 kV X-ray  & A3AC2   & 30.0 & 2.040  & 0.045 & $\sim$ 1.0 \\
$\gamma$ Clinical (6 MV) & UOW429 &  0.0 & 0.049 & 0.245  & $\sim$ 1.5 \\
e$^{-}$ 6 MeV Clinical  & UOW429 & 0.0 & 0.065  &  0.325 & $\sim$ 1.5 \\
Protons 3 MeV      & PAD$\_$V2 & 20.0  & 0.025 & 0.400 & $\sim$ 5  \\
Protons 3 MeV      & PAD$\_$V4 & 20.0  & 0.030 & 0.330 & $\sim$ 5  \\
\botrule 
\end{tabular}
\end{center}
\label{tab:sensitivities}
\end{table}

From the collected data we could draw some conclusions:
\begin{itemize}
    \item both CSC and p-i-n devices grant a sufficient sensitivity for radiation measurement ( $\gt$ tens of pC cGy$^{-1}$) coupled with a low noise ( $\lt$ 10 pA);
    \item  the potential bias is an important parameter to increase the sensitivity, as expected. 
\end{itemize}

\section{Estimate of detection limits for a given a-Si:H device  \label{estimate_detect}}

In this paragraph, we  consider the performance of a realistic a-Si:H device to estimate the minimum detectable flux of different ionizing radiation types associated with solar flares to obtain  a signal-to-noise ratio (S/N) ratio $\ge$ 5, in the first instance. 
\par
For this exercise we use a "virtual"  device with  characteristics similar to those reported in Table \ref{table:types}: 4x4 mm$^{2}$ area, 10 $\mu$m thickness and 5 pA current noise. The geometrical factor of such a device is 0.5 cm$^2$ sr. To increase both geometrical factor and sensitivity, we may choose, for instance, to build a single, large  device  with an area 25 times larger than a single device, 20x20 mm$^{2}$, or to arrange a bi-dimensional array of 25, 4x4 mm$^{2}$ devices read out in parallel. In the following, we indicate both these equivalent solutions as H-SPACE. The final option for the H-SPACE  design would depend most likely by the noise measured for the whole system of measurements. \par 
The leakage current of the devices increases with the area, while the current noise increases with the square root of the area. Therefore,
for the evaluation of particle detection thresholds,  we plausibly assume that the noise of the H-SPACE device scales by a factor equal to the square root of the ratio of the area with respect to that of the small device, i.e. by a factor of 5. This leads to a noise of 25 pA and consequently to 125 pA limit for the 5$\sigma$  detection level. Using the measured sensitivity of the device A3AC2, as a reference,  and converting the dose-rate in deposited energy per second, a deposited energy of 105 MeV/s is the lower limit to obtain a S/N $\gt 5$, while a S/N of 1 is reached for $\sim$ 21 MeV/s energy deposition. For the present preliminary design of our instrument we aim at estimating the dose-rate and associated current generated by incident solar energetic particles on the a-Si:H devices. Future developments of our a-Si:H sensors and  electronics may allow us to measure single particles, but this will be matter of future improvements. To convert the current measurements in particle fluxes we will benefit of Monte Carlo simulations and beam tests. Multiple sensors to cover different portions of solid angles may be of great help in space. It is  worthwhile to point out that the sensitivity of the H-SPACE devices  is expected to  improve in the future  during the R\&D campaign  foreseen for the HASPIDE project.  For the present exercise, the charged particle energy loss distribution in the H-SPACE device represents a lower limit with respect to actual energy losses ,  since only particles with normal incidence on the sensor area are considered, condition that strictly applies to  photons only.

In the following, we simulate the H-SPACE device performance using the FLUKA  Monte Carlo program \cite{flukacern1,flukacern2,flair}. For the moment, we have taken into consideration the characteristics of a-Si:H by decreasing by 10\% the density of c-Si and consequently,   the number of electrons produced per unit length and per unit energy that is proportional to the  to the material density \cite{pdb22}. 

\par

\subsection{Photon rate detection limits}
The  current signal generated by  photons incident on the detector  depends strongly on the photon energy distribution  that sets the photon interaction probability in the sensitive material and, due to the small thickness of the a-Si:H layer, depends also on the partial containment of photoelectrons or Compton electrons.  
\par 
Photon fluxes observed during solar flares are mainly populated below 10 keV \cite{grigis04,benz17} (see Fig. \ref{fig:hardsoft}). Consequently, we focus on photons in the energy range 3-20 keV for which the photoelectric effect is the dominant process occurring within the a-Si:H sensitive volume. 

\begin{figure}[h]%
\centering
\includegraphics[width=0.7\textwidth]{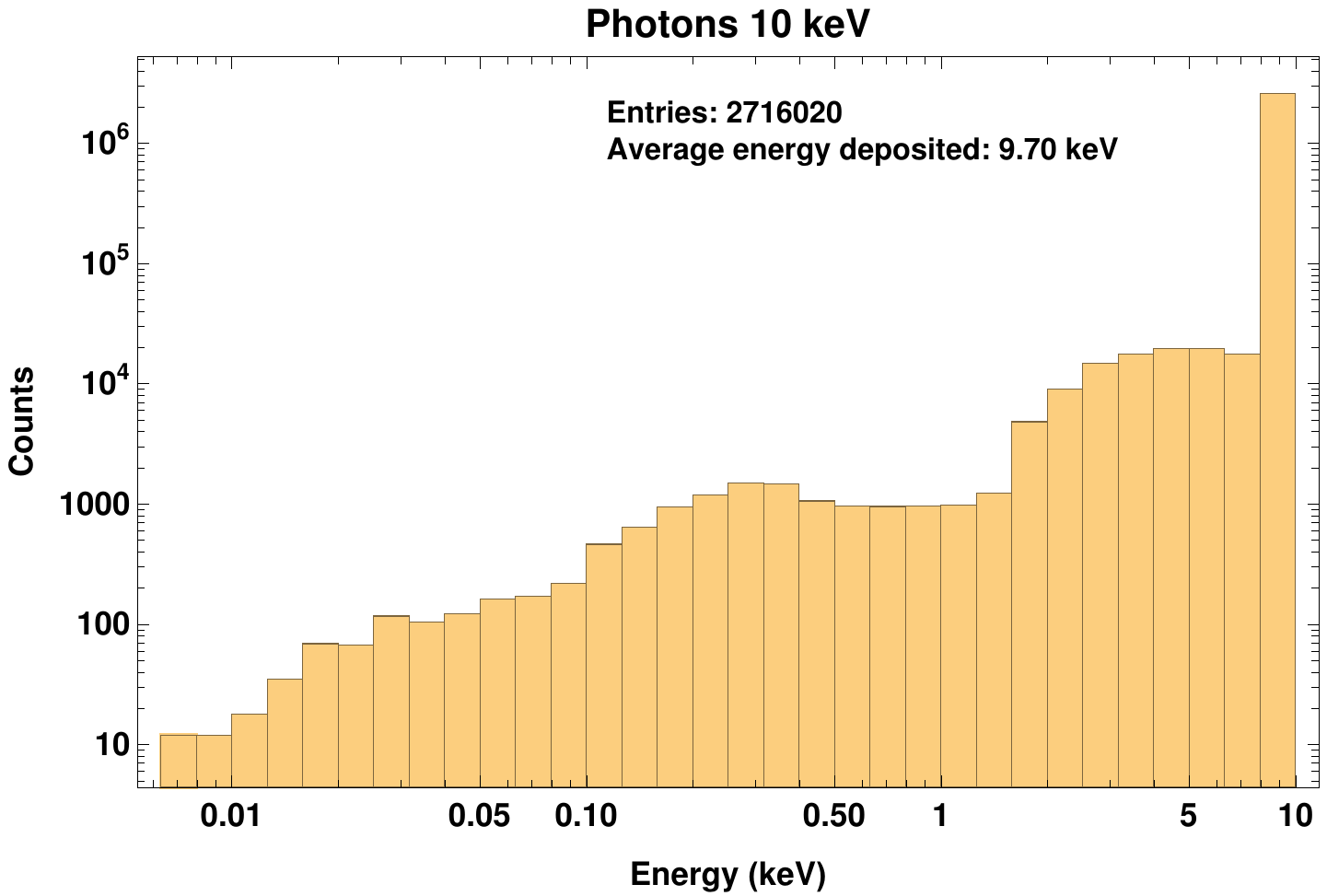}
\caption{FLUKA simulation of energy deposition distribution of 10 keV photons in the H-SPACE device.}
\label{fig:10keVdeposition}
\end{figure}

In Table \ref{table:photons} we have reported the average energy deposition and  the minimum rate of incident photons needed for detection for  several photon energies in the range 3-20 keV.  Since the current measurements are expected to be carried out at 1 Hz, the minimum required energy  deposition per second by photons of given energy is estimated.
It can be observed that, as the photon energy increases, the number of photons needed for a 5$\sigma$ detection increases,  due to decreasing interaction probability of photons within the a-Si:H device and, to a smaller extent, to the increasing difficulty to contain all the energy of the photo-electrons. The percentage of interacting photons are also reported in Table \ref{table:photons}. As an example, Fig. 
\ref{fig:10keVdeposition} shows the  distribution of energies deposited by 10 keV photons traversing the H-SPACE device. Taking the 105 MeV energy deposition threshold set above, 
 the detection limit of 10 keV photons  reaching the device in order to obtain a  S/N ratio of  5  is 1.4 $\cdot$ 10$^5$ s$^{-1}$. 
By considering  these results and by integrating the photon flux reported in Fig. \ref{fig:hardsoft} for the class M flare dated November 9, 2022 between 7 and 16 keV, it is found that the detection limit is met at 4 $\sigma$ in this energy range in the H-SPACE device with the main contribution given by 7-10 keV photons. The reduction of the photon energy range of measurements  to 5 keV, would obviously improve the detection capability of this or stronger (class X) solar flares.


 \begin{table}[h]
 \begin{center}
\caption{ {Rate} detection limits at 5$\sigma$ for  monochromatic photon fluxes.} \label{table:photons}%
\begin{tabular}{cccc}
\toprule
 Energy$_{\gamma}$  & Energy$_{deposited}$ & Interacting photons& Minimum  photon rate detectable\\
 (keV) & (keV)  & (\%) & with HASPIDE-SPACE (photons/s)\\
\midrule
  3.0  & 3.0 $\pm$ 0.3 & 0.87 &4.0$\cdot$10$^{4}$  \\
  5.0  & 4.9 $\pm$ 0.4 & 0.40 &5.4$\cdot$10$^{4}$  \\
  7.0  & 6.9 $\pm$ 0.6 & 0.18 &8.5$\cdot$10$^{4}$  \\
  8.0  & 7.8 $\pm$ 0.8 &  0.13&1.0$\cdot$10$^{5}$  \\
 10.0  & 9.7 $\pm$ 1.3 & 0.08 &1.4$\cdot$10$^{5}$  \\
 15.0  & 14.0 $\pm$ 3.1 & 0.04&3.8$\cdot$10$^{5}$  \\
 20.0  & 17.5 $\pm$ 5.6 & 0.009&6.7$\cdot$10$^{5}$  \\
\botrule
\end{tabular}
\end{center}
\end{table}

\subsection{Electron rate detection limits}
\begin{figure}[h]%
\centering
\includegraphics[width=0.7\textwidth]{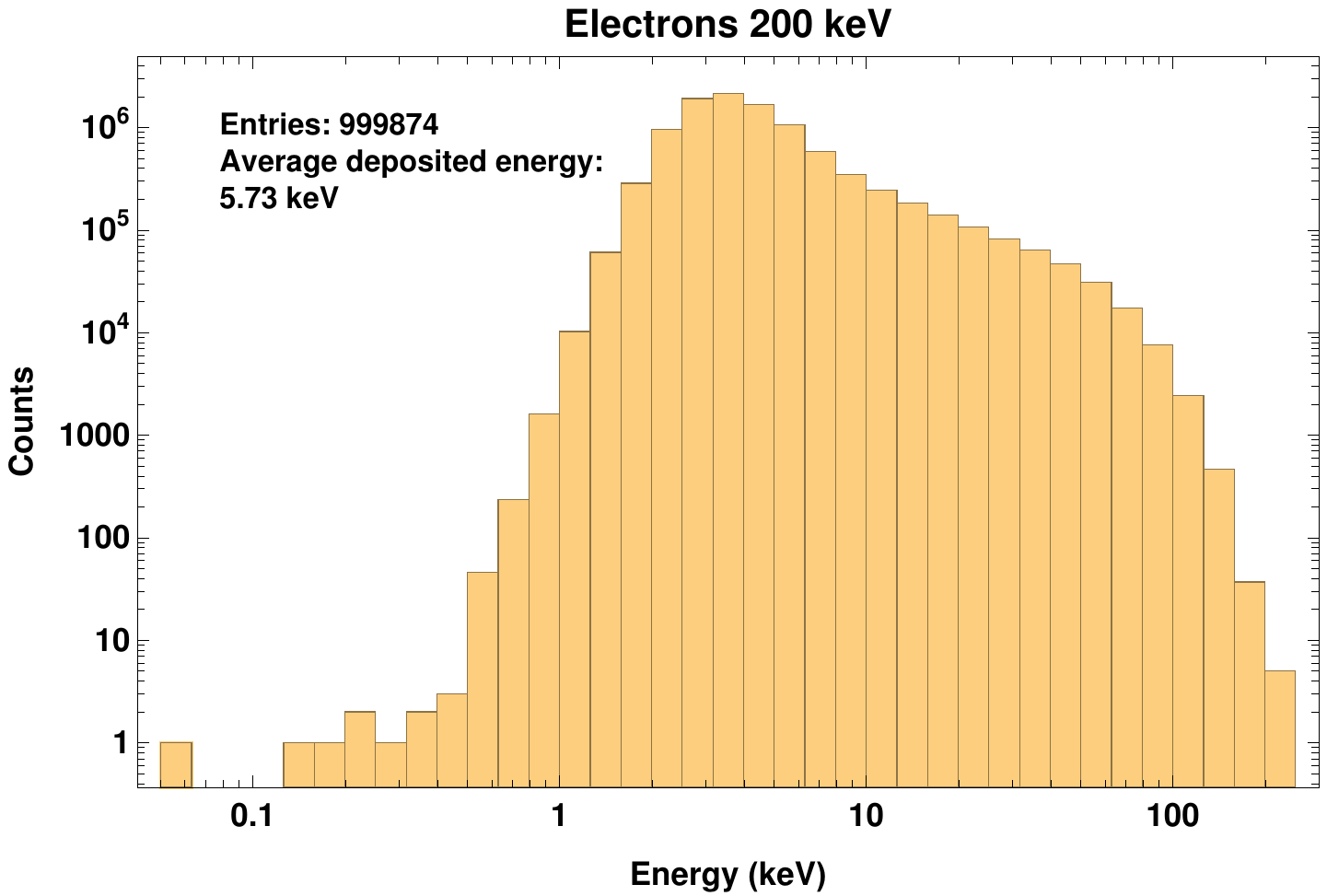}
\caption{FLUKA simulation of energy losses of 200 keV electrons in H-SPACE device.}
\label{fig:200keVdeposition}
\end{figure}

Experimental results on electron energy deposits in thin silicon layers are reported in \cite{2011Meroli,2017Dourki}. 
Table \ref{table:electrons} summarizes the simulation results for deposited energy and minimum number of detectable electrons  per second incident perpendicularly on the H-SPACE sensor surface. 
Looking at the solar electron energy spectra down to 200 keV reported in Fig. \ref{fig:ele_sol} (for energies below 200 keV see for instance \cite{Dresing_2020}), we simulate the 50-1000 keV energy range because higher energy particles give a negligible contribution to the current signal. As an example, Fig. \ref{fig:200keVdeposition} shows the deposited energy distribution obtained for 200 keV electrons. The average is 5.7 keV despite the distribution is populated up to 200 keV. 

 \begin{table}[h]
 \begin{center}
\caption{Five $\sigma$ detection limits  for  monochromatic electron fluxes.} \label{table:electrons}%
\begin{tabular}{ccc}
\toprule
 Energy  & Deposited Energy & Minimum  electron rate detectable \\
 (keV) & (keV)  &  with HASPIDE-SPACE (electrons/s)\\
\midrule
  50.0  & 28.9 $\pm$ 16.0 & 3.6$\cdot$10$^{3}$  \\
  100.0  & 12.1 $\pm$ 13.4 &  8.7$\cdot$10$^{3}$  \\
  200.0  & 5.7 $\pm$ 7.3 &  1.8$\cdot$10$^{4}$  \\
  250.0  &  4.7 $\pm$ 5.6 &  2.2$\cdot$10$^{4}$  \\
 300.0  & 4.1 $\pm$ 4.7 & 2.6$\cdot$10$^{4}$  \\
 500.0  & 3.3 $\pm$ 3.7 & 3.2$\cdot$10$^{4}$  \\
 1000.0  & 2.8 $\pm$ 3.3 & 3.8$\cdot$10$^{4}$  \\
\botrule
\end{tabular}
\end{center}
\end{table}

The simulation results  clearly show that the minimum electron  rate  required for detection can be obtained by measuring low-energy electrons that deposit large amounts of energy in the device. 
By integrating the electron flux in the range 200-250 keV and by taking into account the H-SPACE geometrical factor, we obtain $\sim 1450$ electrons/s. This result is unsatisfactory for solar electron detection  being the S/N ratio $\sim 0.1 $. However, in \cite{Dresing_2020} it is shown that below 100 keV the solar electron energy spectrum  maintains a power-law or a broken power-law trend with respect to higher energies. In principle, by extending the electron measurements down to 50 keV,  a factor of more than 2.5 in the deposited energy and more than a factor of 5 in particle flux would be gained. We will evaluate,  by comparing Monte Carlo simulations and beam-test results, how the S/N ratio for electron detection may increase with respect to the background of other particles.  

\subsection{Proton  rate detection limits}
\begin{figure}[h]%
\centering
\includegraphics[width=0.7\textwidth]{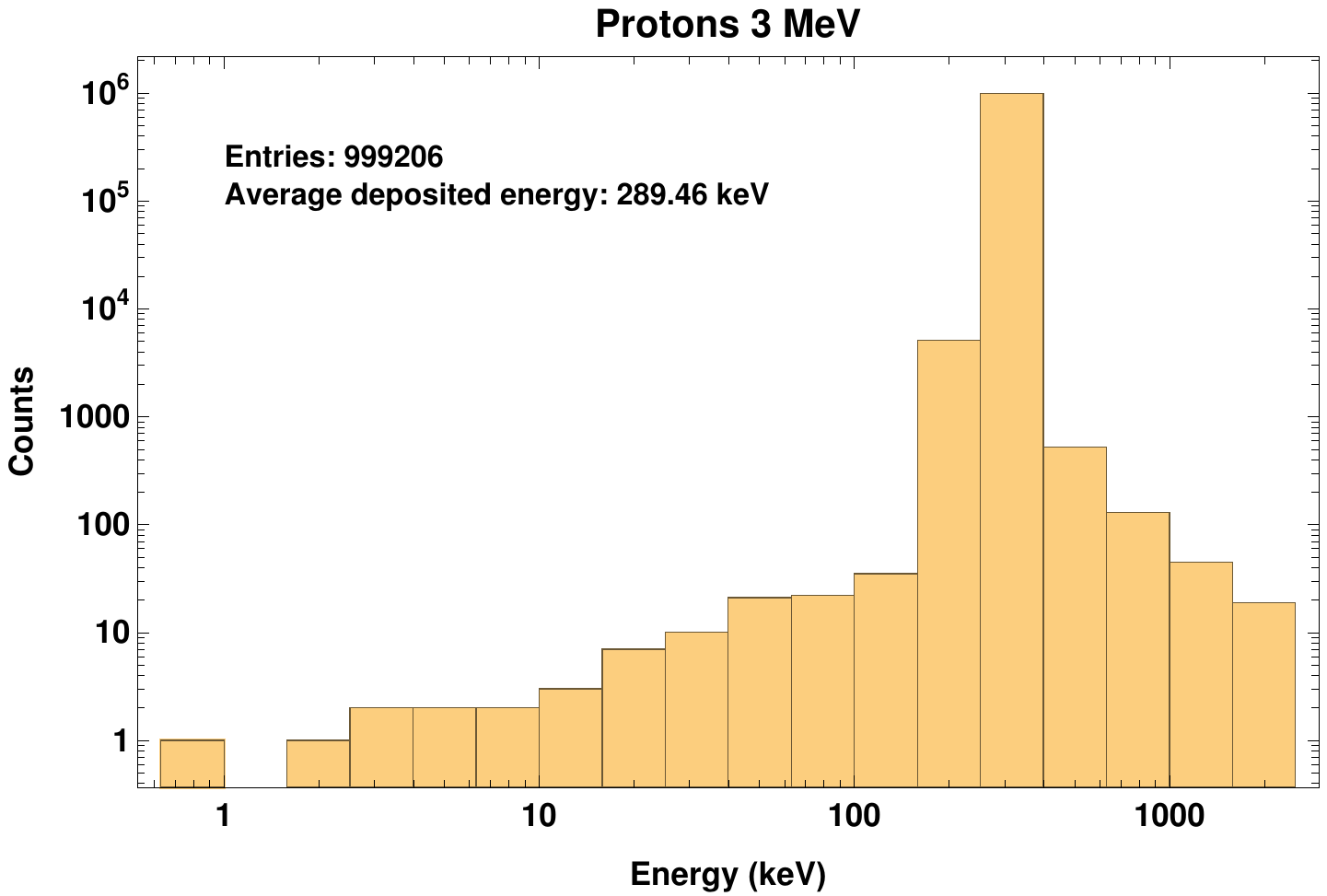}
\caption{FLUKA simulations of energy losses of 3 MeV protons in the H-SPACE device.}
\label{fig:3MeV_proton_deposition}
\end{figure}

\begin{figure}[h]%
\centering
\includegraphics[width=0.7\textwidth]{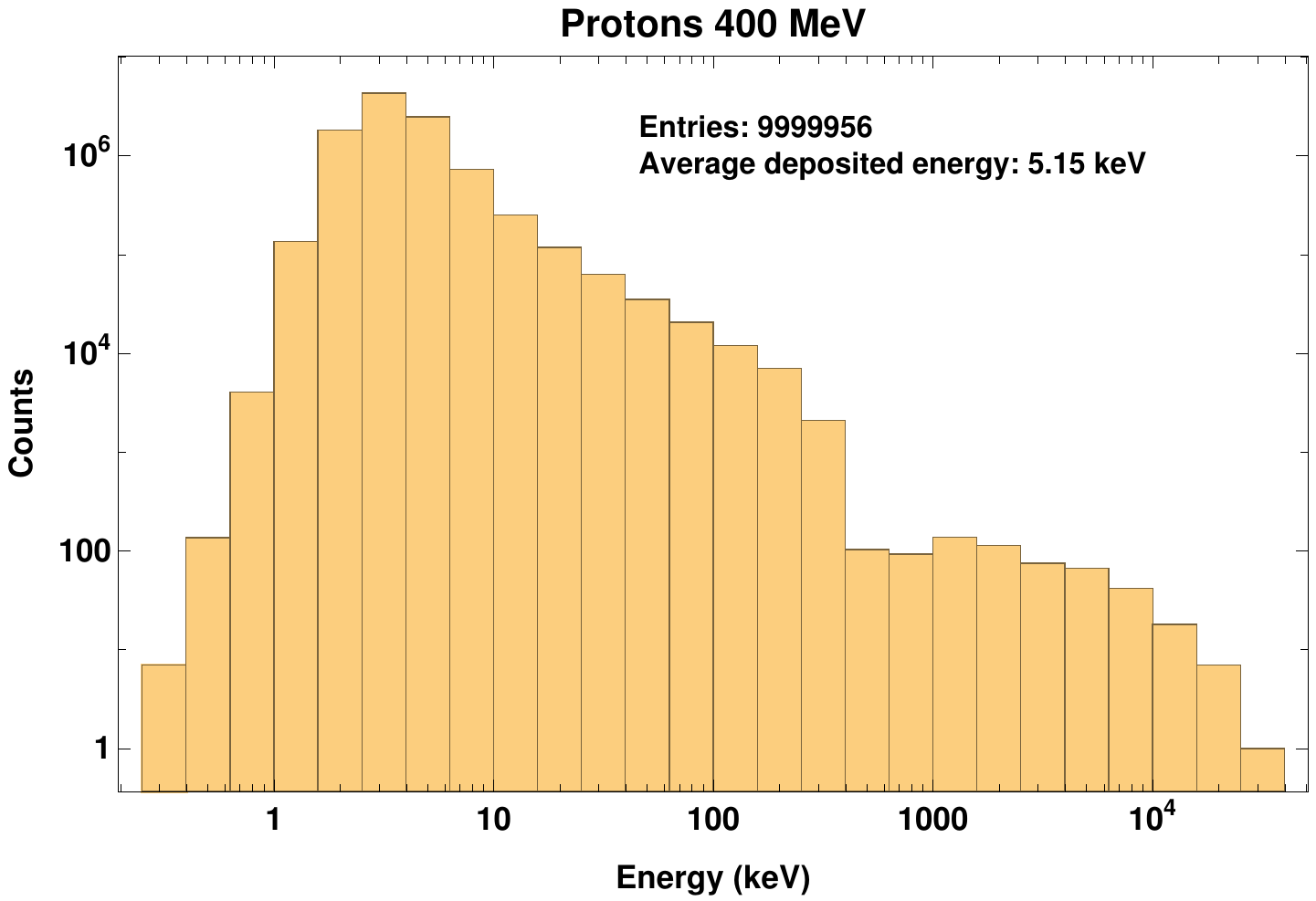}
\caption{FLUKA simulation of energy losses of 400 MeV protons in H-SPACE device.}
\label{fig:400MeV_proton_deposition}
\end{figure}

In Table \ref{table:protons} we have reported the average values of the energy losses   of protons of different energies incident perpendicularly on the H-SPACE sensitive area. The minimum proton  rate   needed for detection with a S/N ratio of 5 is also reported. As in the case of photons and electrons, Table \ref{table:protons} shows how the minimum detectable  proton rate  with S/N $\ge$ 5  varies with proton energy.
For our proposal it is important to have the possibility to monitor solar protons  between 3 MeV (Fig. \ref{fig:3MeV_proton_deposition}) and 400 MeV  (Fig. \ref{fig:400MeV_proton_deposition}).

The average energy deposits vary between  5.1 keV and 289.5 keV. 
This means that the needed number of protons/s crossing the H-SPACE device to have a S/N = 5 scales by the same factor.

\begin{table}[h]
 \begin{center}
\caption{ Rate detection limits  of  monochromatic protons   at 5$\sigma$.} \label{table:protons}%
\begin{tabular}{cccc}
\toprule
 Energy  & Deposited Energy & Minimum  proton rate detectable\\
 (MeV) & (keV)  &  with HASPIDE-SPACE (protons/s)\\\midrule
  3.0  & 289.5 $\pm$ 19.5 &   3.6$\cdot$10$^{2}$   \\
  5.0  & 141.5 $\pm$ 14.5 &   7.4$\cdot$10$^{2}$   \\
 10.0  &  75.6 $\pm$ 14.5 &  1.4$\cdot$10$^{3}$   \\
 20.0  &  43.1 $\pm$ 24.9 &  2.4$\cdot$10$^{3}$   \\
 50.0  &  21.0 $\pm$ 34.5 &  5.0$\cdot$10$^{3}$   \\
 70.0  &  16.3 $\pm$ 35.1 &  6.4$\cdot$10$^{3}$   \\
100.0  &  12.4 $\pm$ 30.7 &  8.5$\cdot$10$^{3}$   \\
200.0  &   7.6 $\pm$ 30.8 & 1.4$\cdot$10$^{4}$   \\
400.0  &   5.1 $\pm$ 35.2 & 2.1$\cdot$10$^{4}$   \\
600.0  &   4.3 $\pm$ 40.4 & 2.6$\cdot$10$^{4}$   \\
\botrule
\end{tabular}
\end{center}
\end{table}

We use Table \ref{table:protons} to evaluate the detection capability of the H-SPACE sensor for the SEP events reported in Table \ref{tab2} and Figs. \ref{fig:sep_onset}-\ref{fig:fit_peak_n}: 
\begin{itemize}
\item February 23 1956: Fig. \ref{fig:sep_onset} shows the measured proton spectrum at the onset; by integrating the spectrum in the 500-700 MeV range of measurement availability, we obtain $\sim $ 224 protons/s across our device at the onset. FLUKA simulations associate to this proton  flux an average energy deposit of 963 keV, i.e. undetectable. Of course in space also lower energy particles would be measured.
\item February 23 1956: Fig. \ref{fig:sep_peak} shows the spectrum at the peak. For the same energy bin as above we obtain 55250 protons/s corresponding to 238 MeV, i.e. a S/N ratio of 2.3 and therefore, within possible detection range. For GCR/SEP discrimination, detection limits of 2.5-3 will still retain the capability to discriminate the SEP dynamics from noise.
\item October 28, 2021: for this event data down to a few MeV are available on the Solar Orbiter archive. Fig. \ref{fig:onset2810} shows the measured proton flux at the onset (black dots). If we consider the energy bin 80-100 MeV where the spectrum is more populated, by integrating the data, we obtain $\sim$ 10 protons/s crossing the H-SPACE device, corresponding to 0.15 MeV/s of deposited energy, i.e. the event would not be detected;
\item October 28, 2021: Fig. \ref{fig:onset2810} shows the measured proton fluxes at the peak (black squares). If we consider again the energy bin 80-100 MeV for the peak, integrating the data, we obtain $\sim$ 142 protons/s, corresponding to 2.0 MeV/s of deposited energy. Also in this case the event would not be detectable.
\item Similar results would be obtained for the December 13, 2006 event, characterized by a similar intensity with respect to that of  the event dated October 28, 2021.

\end{itemize}

In conclusion, with just one layer of the H-SPACE device we would be able to detect SEP proton fluxes above hundreds of MeV only at the  peak of extreme, rare events with a S/N ratio $>$ 3.  

\section{A tentative demonstrator and measuring strategy}\label{demonstrator}
 Our future work will be aimed at showing that a  demonstrator consisting of active and passive layers of sensors and material, respectively, may be built to monitor solar flares and the evolution of medium-strong SEP events in space. 
The separation of different particle species and photons and the attempt to measure the differential flux of protons, will critically depend on the timing of the arrival of photons, electrons and protons at the instrument. X rays will arrive first  and will stop in the first plane of sensors. The superposition of arrival of relativistic electrons and protons will follow, but the current that we will measure in each sensor plane  will be different for electrons and protons due to the different penetration characteristics of the two kind of particles through the layers of passive material of various thickness. In the first plane we will observe electrons and protons. Deep in the detector, we will measure the current associated with the passage of protons only through a varying number of sensitive planes, depending on particle's energy. At present time, we would not be able to follow the onset of even strong SEP events, but we should at the peak when the particle energies decrease. To improve the instrument performance  we will count on a very long campaign of beam tests with photons in the keV range, electrons in the MeV range and protons of different energies.
The experiment  data will be reconciled with Monte Carlo simulations. 
We will simulate  the performance of our instrument from flare detection through the arrival of protons on the basis of real event observations. Basically, following the approach described in \cite{klein22}  for the October 28, 2021 event.


\subsection{Soft X-rays} 
A device capable of photon detection at 4 $\sigma$ for energies above 5 keV could be built given the available deposition techniques of a-Si:H on kapton substrate \cite{2023Menichellitubino} by considering also  a deposition of a metal layer of hundreds of nm on top the sensors for visible light shielding. 
To  further increase the detection power of the device, we could  use identical sensitive layers to sum up the current signals in coincidence. 

\subsection{Electrons}
For one same event, electrons will arrive  at the sensor typically after tens of minutes with respect to photons, hence, in principle, we could use  the same device used for photons for the detection of these particles down to 50 keV.
Detailed Monte Carlo simulations, validated by test-beam measurements, will be carried out to estimate precisely the current signal needed for electron detection as a function of the thickness of the material to be placed in front of the sensitive a-Si:H region and of the thickness of the substrate (300 nm may be feasible). 

\subsection{Protons} 
Special care will be needed to measure the energy spectra of solar protons. In section \ref{3.proton}, we have pointed out that a detector able to measure solar protons up to 400 MeV during medium-strong events will allow us to reliably estimate the particle differential flux up to GeV energies above the background of GCRs. 
From table \ref{table:protons} we observe that by reducing the proton energy in the instrument we will improve sensibly the S/N ratio. The technique of proton energy degradation using  layers of passive material is well known and applied in several domains. Hence we propose to use  layers  of passive material (typically tungsten\footnote{\url{https://www.nist.gov/pml/stopping-power-range-tables-electrons-protons-and-helium-ions}}) to obtain two results at once: to degrade the energy of the protons and absorb the protons ranged down in energy, in order to be able to measure the proton differential flux up to 400 MeV energy. 
Since 50 keV electrons and hundreds of MeV protons have similar speeds, these particles  will reach the instrument at the same time. However, because of the very different capability of propagation  in the passive material of these electrons and protons, we will be able to separate the two different particle species.  

To estimate the proton fluxes, the current measurements will be converted in number of protons penetrating different layers of passive material. For this achievement, we will use both Monte Carlo simulations and beam tests. In particular, due to the varying spatial distribution of solar particles during the evolution of the SEP events and to the particle interactions in the passive material, the instrument will be tested at different angles of incidence of the beam. Adding single sensors in space to cover the solid angle (with zero bias) around the main instrument may also help.
The   instrument should contain approximately 8 cm of tungsten to stop 75\% of 400 MeV protons incident perpendicularly. The typical weight and power consumption for a radiation monitor are 1 kg and 1 W, respectively \cite{ngrm}. In order to limit the role of particle crossing the instrument on the side, we need to use extra passive material surrounding the sensitive part of the instrument and consequently we may need to drop the strict requirement of 1 kg on the weight  while the power consumption may remain competitive by limiting the power bias to a few V. The actual limits will be set on the basis of the constraints of the possibly assigned space mission. The number of sensitive planes mounted between  slabs of tungsten of different thickness and passive material surrounding the a-Si:H sensors will be optimized with future Monte Carlo simulations.

\subsection{a-Si:H device improvements}
A boost for the detection capability of sensitive layers of our instrument could come from the R\&D program presently in progress. We may evaluate to: 

\begin{itemize}
\item increase the sensor surface: we expect a linear increase of current signal and a noise increasing with the square root of the device area; 
\item increase the sensor thickness;
a thickness of 30 $\mu$m is possible, for a gain factor of $\sim$ 3; 
\item noise reduction below 1 pA: we are currently developing a readout chip in 28 nm technology; this is expected to reduce the noise by a factor 2-5;
\item increase sensitivity: we are experimenting different contact types and also different deposition techniques with a possible gain factor of $\sim$ 5.
\end{itemize}


Finally, the optimization of the detector design, the estimate of the final geometrical factor, the number of detector units to be adopted should be assessed on the basis of the aims of the mission hosting the instrument. 
In case the instrument would be aimed at monitoring extreme SEP events (fluence larger than 10$^8$ protons/cm$^2$), one unit with a first special layer to detect X-ray and electrons followed by 4 sensitive and 4 passive layers for a total of not less than 8 cm thickness; 
to monitor the evolution of strong to medium intensity events (similar to the event dated December 13, 2006), several units could be considered to increase the geometrical factor by not less than a factor of 40. Finally,
at least  four instruments observing different solid angles with respect to spacecraft-Sun direction should be considered to study the SEP pitch angle. 

\section{Conclusions}\label{sec13}
In this work we have reported a preliminary study of an instrument based on a-Si:H as active material for solar photon, electron and proton  monitoring.  We have demonstrated that proton measurements up to 400 MeV range allow us to reasonably extend the  parameterization of SEP data above 1 GeV which is the energy range of major interest for Space Weather. At the moment this is feasible at the peak of extreme events only. Future, dedicated Monte Carlo simulations and R\&D activities will lead to an optimization of the detector design.  This preliminary investigation represents the  basis of an effort to respond to future space agencies calls  for the development of innovative space-based instruments.



\bmhead{Acknowledgments}
We are grateful to the anonymous referee for his/her very constructive comments that allowed us to substantially improve our manuscript.\\
The HASPIDE project is funded by INFN through the CSN5 and was partially supported by the
“Fondazione Cassa di Risparmio di Perugia” RISAI project n. 2019.0245.
F. Peverini has a PhD scholarship funded by the PON program.
\section*{Declarations}

\begin{itemize}
\item Funding 
The HASPIDE project is funded by INFN through the CSN5 and was partially supported by the
“Fondazione Cassa di Risparmio di Perugia” RISAI project n. 2019.0245.
\item Conflict of interest/Competing interests\\
The authors declare that they have no conflict of interest.
\item Ethics approval 
 'Not applicable'
\item Consent to participate
 'Not applicable'
\item Consent for publication
 'Not applicable'
\item Availability of data and materials\\
 All data mentioned in the paper are publicly available.
\item Code availability \\
 Code is available upon request
\item Authors' contributions\\
Conceptualization: Catia Grimani, Leonello Servoli; Methodology: Leonello Servoli, Mattia Villani, Federico Sabbatini, Catia Grimani, Lucio Calcagnile, Anna Paola Caricato, Maurizio Martino, Giuseppe Maruccio, Anna Grazia Monteduro, Gianluca Quarta, Silvia Rizzato; Formal analysis and investigation: Leonello Servoli, Catia Grimani, Mattia Villani, Federico Sabbatini, Roberto Catalano, Giuseppe Antonio Pablo Cirrone, Tommaso Croci, Giacomo Cuttone, Luca Frontini, Maria Ionica, Keida Kanxheri, Matthew Large, Valentino Liberali, Giovanni Mazza, Mauro Menichelli, Arianna Morozzi, Francesco Moscatelli, Stefania Pallotta, Daniele Passeri, Maddalena Pedio, Marco Petasecca, Giada Petringa, Francesca Peverini, Lorenzo Piccolo, Pisana Placidi, Alberto Stabile, Cinzia Talamonti, James Richard Wheadon; Writing - original draft preparation: Catia Grimani, Leonello Servoli; Writing - review and editing: Mattia Villani, Federico Sabbatini, Michele Fabi, Lucio Calcagnile, Anna Paola Caricato, Roberto Catalano, Giuseppe Antonio Pablo Cirrone, Tommaso Croci, Giacomo Cuttone, Sylvain Dunand, Luca Frontini, Maria Ionica, Keida Kanxheri, Matthew Large, Valentino Liberali, Maurizio Martino, Giuseppe Maruccio, Giovanni Mazza, Mauro Menichelli, Anna Grazia Monteduro, Arianna Morozzi, Francesco Moscatelli, Stefania Pallotta, Daniele Passeri, Maddalena Pedio, Marco Petasecca, Giada Petringa, Francesca Peverini, Lorenzo Piccolo, Pisana Placidi, Gianluca Quarta, Silvia Rizzato, Alberto Stabile, Cinzia Talamonti, James Richard Wheadon, Nicolas Wyrsch; Funding acquisition: Leonello Servoli; Resources: ; Supervision: Catia Grimani, Leonello Servoli.
\end{itemize}

\bibliography{sn_ASI_paper_rev3}


\end{document}